\documentclass[manuscript,prb,aps,superscriptaddress,floatfix]{revtex4} 
\usepackage{graphicx,amssymb,amsmath} 
\begin{document} 

\title{A Molecular Dynamics Computer Simulation Study of Room-Temperature Ionic Liquids:\\
II.\ Equilibrium and Nonequilibrium Solvation Dynamics}

\author{Y. Shim} 
\affiliation{Department of Physics, Seoul National University, Seoul 151-747, Korea} 
\affiliation{Department of Chemistry, Carnegie Mellon University, Pittsburgh, PA 15213, U.S.A.} 

\author{M. Y. Choi} 
\affiliation{Department of Physics, Seoul National University, Seoul 151-747, Korea} 
\affiliation{Korea Institute for Advanced Study, Seoul 130-722, Korea} 

\author{Hyung J. Kim} 
\altaffiliation{Author to whom correspondence should be addressed. E-mail: hjkim@cmu.edu} 
\affiliation{Department of Chemistry, Carnegie Mellon University, Pittsburgh, PA 15213, U.S.A.} 

\begin{abstract}
The molecular dynamics (MD) simulation study of solvation structure and free energetics in
1-ethyl-3-methylimidazolium chloride and 1-ethyl-3-methylimidazolium hexafluorophosphate using a probe solute in the preceding article [Y.\ Shim, M.\ Y.\ Choi and H.\ J.\ Kim, J.\ Chem.\ Phys. xxxxxxxxxx] is extended to investigate dynamic properties of these liquids.  Solvent fluctuation dynamics near equilibrium are studied via MD and associated time-depedent friction is analyzed via the generalized Langevin equation.  Nonequilibrium solvent relaxation following an instantaneous change in the solute charge distribution and accompanying solvent structure reorganization are also investigated.  Both equilibrium and nonequilibrium solvation dynamics are characterized by at least two vastly different time scales---a subpicosecond inertial regime followed by a slow diffusive regime.  Solvent regions contributing to the subpicosecond nonequilibrium relaxation are found to vary significantly with initial solvation configurations, especially near the solute.  If the solvent density near the solute is sufficiently high at the outset of the relaxation, subpicosecond dynamics are mainly governed by the motions of a few ions close to the solute.  By contrast, in the case of a low local density, solvent ions located not only close to but also relatively far from the solute participate in the subpicosecond relaxation.  Despite this difference, linear response holds reasonably well in both ionic liquids.

\end{abstract}

\maketitle

\section{Introduction}

In the preceding paper,\cite{rtil:shim1} referred to as I hereafter, we have investigated the equilibrium solvation structure and free energetics in two different room-temperature ionic liquids (RTILs), 1-ethyl-3-methylimidazolium chloride ($\mathrm {EMI^+Cl^-}$) and 1-ethyl-3-methylimidazolium hexafluorophosphate ($\mathrm{EMI^+PF_6^-}$) by employing a diatomic and a benzenelike probe solute.\cite{rtil:shim1} We have found that these solvents show a dramatic electrostrictive behavior; i.e., the radial and angular distributions of solvent ions around the solute become structured markedly as the solute charge separation is increased.  This results in, e.g., significant outer-sphere reorganization free energies, which are comparable to and sometimes in excess of those in highly dipolar solvents.\cite{rtil:shim1,shim}  Also the effective polarity of these RTILs, which measures their solvating power of dipolar solutes, is very high. 
The free energy curves relevant to solvent fluctuations and associated local force constants vary with the solute charge distributions.  This indicates that a linear solvation scheme is generally not valid for RTILs. Nevertheless the harmonic approximation with state-dependent force constants provides a reasonable framework to quantify, e.g., the solvent reorganization free energy.

In this article, we continue our investigation of solvation in RTILs, focusing on their dynamic properties.  We study equilibrium solvation dynamics and examine their variations with the solute electronic structure.  Using the generalized Langevin equation (GLE) description, we analyze time-dependent solvent friction.  As for nonequilibrium, time-dependent Stokes shifts subsequent to an instantaneous change in the solute charge distribution and accompanying solvent structure relaxation are investigated.  Underlying molecular motions and their roles in nonequilibrium relaxation are also examined.
Comparison is made between equilibrium and nonequilibrium solvation dynamics to gain insight into linear response in RTILs.

The outline of this paper is as follows: In Sec.~II we briefly review the time-correlation function and GLE descriptions of solvation dynamics. 
The molecular models and simulation methods employed in our study are summarized in Sec.~III. In Sec.~IV we present an analysis of the simulation results for equilibrium fluctuation and nonequilibrium relaxation dynamics in EMI$^+$Cl$^-$ and EMI$^+$PF$_6^-$.  Concluding remarks are offered in Sec.~V.

\section{Solvation Dynamics and Generalized Langevin Equation Description}
\label{sec:theory}

Here we review briefly the time-correlation function description of solvation dynamics to establish our notation.  We also give a brief account of the GLE equation approach to solvation.  

As in I,  the solute is assumed to be described by two nonpolarizable electronic states, $a$ and $b$.  For a given solvent configuration ${\cal Q}$ and a solute electronic state $i$ ($=a,b$), we denote the total energy of the combined solute-solvent system as $E_i({\cal Q})$. 
The Franck-Condon (FC) energy $\Delta E_{a\to b}({\cal Q})$ associated with the $a\to b$ transition in the presence of ${\cal Q}$ is [Eq.~(4) of I]
\begin{equation}
\Delta E_{a\to b}({\cal Q})= E_a({\cal Q}) - E_b({\cal Q})\ .
\label{eq:egap}
\end{equation}
As mentioned in I, $\Delta E_{a\to b}$ provides a very convenient variable to describe the collective solvent influence on the solute.\cite{warshel} 
From experimental perspective, by measuring the solute absorption or fluorescence band, in particular, modulations in $\Delta E_{a\to b}$ induced by solvation, we can probe solvent structure and dynamics in the presence of the solute in state $a$.  Henceforth, we will refer to the originating state of the probe FC transitions (viz., the state $a$ in $\Delta E_{a\to b}$) as the active state, while the other electronic state involved in the FC transitions (i.e., $b$ in $\Delta E_{a\to b}$) as the reference state.
In the remainder of this paper, we focus mainly on dynamics associated with $\Delta E_{a\to b}$ under both equilibrium and nonequilibrium solvation conditions.

We first consider the equilibrium solvation situation.  
Fluctuating dynamics of $\Delta E_{a\to b}$ near equilibrium in the presence of a solute in state $a$ is usually described by the normalized time-correlation function
\begin{equation}
C_{a/b}(t) \equiv \frac{\langle \delta \Delta E_{a\to b}(0)\, \delta \Delta E_{a\to b}(t)\rangle}
                        {\langle (\delta \Delta E_{a\to b})^2\rangle}
\label{eq:eq:tcf}
\end{equation}
with 
\begin{equation}
\langle \delta \Delta E_{a\to b}(0)\, \delta \Delta E_{a\to b}(t)\rangle
= \int \! d{\cal Q}_0\, f_a^{\text{eq}}({\cal Q}_0) \,
\delta \Delta E_{a\to b}({\cal Q}_0)\, \delta \Delta E_{a\to b}({\cal Q}_t)\ ,
\label{eq:eq:av}
\end{equation}
where $\langle \ldots \rangle$ denotes an average with the ensemble distribution function $f_a^{\text{eq}}$ in equilibrium with the $a$-state solute, $\delta \Delta E_{a\to b}\equiv \Delta E_{a\to b}-\langle\Delta E_{a\to b}\rangle$ is the deviation of $\Delta E_{a\to b}$ from its equilibrium average, and ${\cal Q}_0$ and ${\cal Q}_t$ represent, respectively, solvent configurations at time 0 and $t$.  We point out that for the equilibrium ensemble average in Eq.~(\ref{eq:eq:tcf}), the solute state which determines the equilibrium ensemble distribution at $t=0$ and the active state which the FC transitions to monitor solvent dynamics originate from are identical. 

The normalized time-dependent Stokes shift function
\begin{equation}
S_{a/b}(t)\equiv\frac{\overline{\Delta E_{a\to b}}(t)
-\overline{\Delta E_{a\to b}}(\infty)}
{\overline{\Delta E_{a\to b}}(0)
-\overline{\Delta E_{a\to b}}(\infty)}\ ,
\label{eq:stokes:response}
\end{equation}
directly accessible via time-dependent fluorescence,\cite{ref3,ref4} is widely used to describe nonequilibrium solvation dynamics.  In Eq.~(\ref{eq:stokes:response}),
$\overline{\Delta E_{a\to b}}(t)$ denotes the average FC energy associated with the $a\to b$ transition at time $t$ after an instantaneous change in the solute electronic structure from the charge distribution of the reference state $b$ to that of the active state $a$.  The average is taken over the initial distribution of solvent configurations in equilibrium with the reference $b$-state\cite{comment:noneq:relax}
\begin{equation}
\overline{\Delta E_{a\to b}}(t) = \int \! d{\cal Q}_0\, f_b^{\text{eq}}({\cal Q}_0) \,
 \Delta E_{a\to b}({\cal Q}_t)\ .
\label{eq:stokes:gap}
\end{equation}
From Eqs.\ (\ref{eq:egap}) and~(\ref{eq:stokes:gap}), equilibrium and nonequilibrium averages satisfy
\begin{equation}
\overline{\Delta E_{a\to b}}(0) = -\langle \Delta E_{b\to a} \rangle\ =
-\overline{\Delta E_{b\to a}}(\infty)\ .
\label{eq:ave:relation}
\end{equation}
We stress that for both nonequilibrium $S_{a/b}(t)$ and equilibrium $C_{a/b}(t)$, solvent dynamics occur in the presence of the active $a$-state solute.  The major difference between the two lies in the character of their ensembles.  In the linear response regime, $S_{a/b}(t)$ and $C_{a/b}(t)$ are characterized by the same dynamic behavior.  Its validity in RTILs is investigated in Sec.~\ref{subsec:noneq} below.

We turn to a stochastic description of solvation dynamics using a generalized Langevin equation (GLE).  For brevity, the subscripts representing the solute electronic states are suppressed in the remainder of this section.
The Mori-Zwanzig projection\cite{gle} on to a set of dynamical variables
$\{\delta\Delta E, \delta\Delta \dot{E}\}$ yields 
\begin{equation}
\delta\Delta \ddot{E} (t)=-\omega_s^2\ \delta \Delta E (t)-\int_0^t d{t^\prime}\ \zeta(t-t^\prime) \,\delta\Delta \dot{E} (t^\prime) + R (t)\ ,
\label{eq:gle}
\end{equation}
where the overdot denotes the time derivative, $\zeta (t)$ and $R (t)$ the time-dependent friction and random force (scaled by the inertia associated with $\delta\Delta E$ motions), respectively, and $\omega_s$ the solvent frequency  defined as
\begin{equation}
\omega_s^2 \equiv \frac{\langle(\delta\Delta \dot{E})^2\rangle}{\langle(\delta\Delta E)^2\rangle}\ .
\label{eq:solv:freq}
\end{equation}
The friction and the random force are related via the so-called second fluctuation-dissipation theorem:
\begin{equation}
\zeta (t) = \frac{\langle R(0)R(t)\rangle}{\langle(\delta\Delta \dot{E})^2\rangle}\ .
\label{eq:fric}
\end{equation}
$\zeta(t)$ describes dissipative $\Delta E$ motions caused by rapidly-fluctuating $R(t)$, while $\omega_s$ characterizes inertial dynamics of $\Delta E$.\cite{hynes:solvation}
It should be noted that the GLE formulation with Eqs.\ (\ref{eq:gle})--(\ref{eq:fric}) is formally exact. 

It is easy to show that the equilibrium time-correlation function (TCF) in Eq.~(\ref{eq:eq:tcf}) satisfies\cite{gle}
\begin{equation}
\ddot{C} (t)=-\omega_s^2\ C (t)-\int_0^t d{t^\prime}\ \zeta(t-t^\prime) \,\dot{C} (t^\prime) \ .
\label{eq:gle:c}
\end{equation}
With the Fourier-Laplace transform
\begin{equation}
\tilde{C} [z] = \int_0^{\infty} dt\ \textrm{exp}[izt]~C (t)\ ,
\label{eq:laplace}
\end{equation}
we obtain
\begin{equation}
\tilde{C} [z] = \left( -iz+\frac{\omega_s^2}{-iz+\tilde{\zeta}[z]}\right)^{-1}\ ,
\label{eq:cofz}
\end{equation}
which shows that $C(t)$ is governed by the solvent frequecy $\omega_s$ and friction $\zeta(t)$.  For example, near $t=0$, the friction term in Eq.~(\ref{eq:gle:c}) can be ignored, so that  $\ddot{C} (t)\approx -\omega_s^2\ C (t).$
This yields $C(t) \approx \exp[-{1\over 2}\omega_s^2 t^2]$, manifesting that the short-time behavior is governed by inertial dynamics.
We also consider the TCF
\begin{equation}
\Phi(t)= \frac{\langle {\cal F_N}(0) {\cal F_N}(t)\rangle} 
              {\langle {\cal F_N}^2\rangle} \ ,
\label{eq:force:tcf}
\end{equation}
where ${\cal F_N}(t) \equiv \delta\Delta \ddot E(t) + \omega^2\,\delta\Delta E(t)$ represents the ``non-conservative'' force on $\delta \Delta E(t)$, viz., the sum of frictional and random forces ($\langle{\cal F_N}\,\delta\Delta E  \rangle=0$). The friction is related to $C(t)$ and $\Phi(t)$ via
\begin{equation}
\tilde \zeta [z]=\frac{(\omega_s^2-z^2)\tilde{C}[z]+iz}{1+iz\tilde{C}[z]}
=\left(1+ \frac{\strut  iz}{\omega^2-z^2}\,\tilde \Phi [z]\right)^{-1} \tilde \Phi[z]\ .
\label{eq:zeta:phi}
\end{equation}
In Sec.~\ref{subsec:eq} below, we use Eq.~(\ref{eq:zeta:phi}) to determine $\zeta(t)$ from the simulation data.

\section{Simulation Methods}
\label{sec:method}

As in I, the simulation cell is comprised of a single rigid solute molecule immersed in either EMI$^+$Cl$^-$ or EMI$^+$PF$_6^-$ solvent, consisting of 112 pairs of rigid cations and anions.  Two different types of solutes, diatomic and benzenelike molecules, are considered.   For each solute type, we have employed two different charge distributions:  For the diatomic solute, a neutral pair (NP) with no charges and an ion pair (IP) with unit charge separation ($q\!=\!\pm e$) are used, where $e$ is the elementary charge.  As for the benzenelike solute, we consider an axially symmetric charge distribution appropriate for regular ground-state benzene (RB)\cite{benzene} and charge-separated dipolar benzene (DB), which is characterized by one electron transfer from one carbon to another carbon atom in the para position, compared to RB.  Thus the dipole moment difference between the RB and DB charge distributions is 13.5\,D.  For EMI$^+$,  we use the AMBER force field\cite{ref6} 
for the Lennard-Jones (LJ) parameters and the partial charge assignments of Ref.~\onlinecite{ref7}.  The united atom representation is employed for the CH$_2$ and CH$_3$ moieties of the ethyl group as well as for the methyl group (Fig.~1 of I).  Experimental geometry determined by X-ray diffraction\cite{ref8} is used.  For Cl$^-$, we use $\sigma\!=\!4.4$\,\AA\ and $\epsilon/k_{\rm B}\!=\!50.4$\,K with Boltzmann's constant $k_{\text B}$.  PF$_6^-$ is described as a united atom with $\sigma\!=\!5.6$\,\AA\ and $\epsilon/k_{\rm B}\!=\!200$\,K.  
For the details of the solute and solvent model descriptions, the reader is referred to I.


The molecular dynamics (MD) simulations are conducted in the canonical ensemble at $T=400$\,K using the DL$\_$POLY program.\cite{ref5} Equilibrium simulations are carried out with 2\,ns equilibration, followed by a 8\,ns trajectory from which averages are computed.  
For each reference state, the nonequilibrium quantities are computed by averaging over 400 distinct initial configurations generated from its 8\,ns equilibrium trajectory and separated by 20\,ps in time.  These initial configurations are used to  generate nonequilibrium solvation by  changing the solute charges from the reference-state to active-state charge distribution instantaneously {\`a} la Franck-Condon.  The subsequent solvent relaxation to new equilibrium is monitored for 2\,ps in each of the initial configurations. 

\section{MD Results and Discussions}

\label{sec:result}

\subsection{Equilibrium solvation dynamics}
\label{subsec:eq}

The MD results for $C_{a/b}(t)$ are presented in Fig.~\ref{fig:coft} for several values of the solvent density $\rho$.  Regardless of the solute and solvent considered, $C_{a/b}(t)$ is characterized by at least two different dynamics:  ultrafast initial relaxation followed by an extremely slow decay.  In Ref.~\onlinecite{shim}, the latter was attributed to ion transport via diffusion and the former to small-amplitude inertial translations of ions.  There it was also pointed out that the long-time relaxation is not a single-exponential decay but can be fitted reasonably well with a stretched exponential.  Despite their relatively high viscosity, short-time dynamics in both EMI$^+$Cl$^-$ and EMI$^+$PF$_6^-$ fall in a subpicosecond time regime and account for more than 50~\% of the entire $C_{a/b}(t)$ relaxation. 

We notice in Fig.~\ref{fig:coft} that the early dynamics of $C_{a/b}(t)$ tend to become faster with the growing charge separation (i.e., dipole moment) of the solute active state.  For instance, as we change the active state charge distribution from NP to IP (and from RB to DB), the solvent frequency $\omega_s$ [Eq.~(\ref{eq:solv:freq})] which describes the initial relaxation behavior of $C_{a/b}(t)$, increases by by 30--60~\%.  Dipolar acetonitrile shows a similar trend according to the present and previous simulation studies.\cite{maroncelli}
The results in Table~\ref{table:result} show that two factors contribute to this trend -- viz., decrease in $\langle (\delta \Delta E_{a\to b})^2\rangle$ and increase in $\langle (\delta \Delta \dot{E}_{a\to b})^2\rangle$.  In I, the former is ascribed to the enhancement in solvent structural rigidity induced by electrostriction as the solute charge separation grows.  As for the latter behavior, the increase in the electrostatic interaction strength, also discussed in I, plays a central role.  To see this, we rewrite $\langle (\delta \Delta \dot{E}_{a\to b})^2\rangle$ as
\begin{eqnarray}
\langle (\delta \Delta \dot{E}_{a\to b})^2\rangle
&\approx& \langle (\sum_\alpha{\bf v}_\alpha\cdot \nabla_\alpha \Delta {E}_{a\to b})^2\rangle 
= {1\over 3}\sum_\alpha\langle {v_\alpha}^2\rangle \langle(\nabla_\alpha \Delta {E}_{a\to b})^2\rangle \label{eq:solv:vel} \nonumber\\
&=&\sum_\alpha {k_BT\over m_\alpha}\langle(\nabla_\alpha \Delta {E}_{a\to b})^2\rangle
=\sum_\alpha {k_BT\over m_\alpha}\langle (\Delta f_\alpha^{\text{coul}})^2\rangle\ , 
\end{eqnarray}
where $\alpha$ labels solvent ions, $m_\alpha$ and ${\bf v}_\alpha$ are the mass and velocity of ion $\alpha$, and $\Delta f_\alpha^{\text{coul}}$ is the difference in the Coulomb forces on $\alpha$ arising from the $a$- and $b$-state solute charge distributions.  For both $\mathrm{NP\to IP}$ and $\mathrm{IP\to NP}$, $\Delta f_\alpha^{\text{coul}}$ is, up to a sign, just the Coulomb force on $\alpha$ arising from the IP charges because there are no electrostatic interactions between the solvent and the NP charge distribution.  We have, for the sake of simplicity, ignored the cation rotations as well as solute translations and rotations in Eq.~(\ref{eq:solv:vel}).  The local solvent density increase close to the solute with the growing solute charge separation (see Figs.\ 2--4, 6 and~7 of I) enhances the solute-solvent electrostatic interactions, viz., the magnitude of $\Delta f_\alpha^{\text{coul}}$.  This results in the increase of $\langle (\delta \Delta \dot{E}_{a\to b})^2\rangle$ according to Eq.~(\ref{eq:solv:vel}).  For the RTILs and aprotic $\mathrm{CH_3CN}$ studied here, $\langle (\delta \Delta \dot{E}_{a\to b})^2\rangle$ and $\langle (\delta \Delta {E}_{a\to b})^2\rangle$ contribute  in a constructive way to the solvent frequency trend with the solute active state charge separation.
    
Another interesting feature in Fig.~\ref{fig:coft} is that $C_{a/b}(t)$ becomes more oscillatory as the $a$-state becomes more dipolar.  According to our earlier study,\cite{shim} the solvent structure enhancement with the increasing solute dipole is responsible for this trend.  Briefly, enhanced structure yields a tighter potential of mean force for ions, which results in more oscillations with higher frequencies, especially for anions (see Fig.~\ref{fig:coft:comp} below).  In contrast to the short-time oscillations, the long-time relaxation of $C_{a/b}(t)$ becomes slower as the active-state charge separation increases; both the exponent $\beta$ and time constant $\tau_0$ of its stretched exponential fit, i.e., $\exp[-(t/\tau_0)^\beta]$, decrease (see Table~\ref{table:result}).
In Ref.~\onlinecite{shim}, this was also attributed to the potential of mean force, which becomes more rugged with the growing solute charge separation.  This decelerates ion transport and thus the long-time decay of $C_{a/b}(t)$. 

In Fig.~\ref{fig:coft:comp}, we decompose $\Delta E_{a\to b}$ into the cation and anion components and analyze the contributions of their autocorrelations and cross correlation to $C_{a/b}(t)$.  The normalization of these three components are such that the sum of the cation and anion autocorrelations and their cross correlation yields $C_{a/b}(t)$.   Figs.\ \ref{fig:coft:comp}(a) and~\ref{fig:coft:comp}(b) show that  in EMI$^+$Cl$^-$, short-time subpicosecond  solvation dynamics are almost completely determined by anions.\cite{shim}
Also the oscillatory behavior of $C_{a/b}(t)$ for $a=\mathrm{IP}$ arises from Cl$^-$.
Because of their small size, Cl$^-$ ions can approach rather close to the solute.  This makes their Coulombic interactions with the solute very strong.  For instance, the Coulomb interaction energy between 
 Cl$^{-}$ and the (+) site of IP at a separation of 3.5~\AA\ [cf.\ Fig.~2(a) of I] is about $-95$\,kcal/mol.  Therefore, translational motions of Cl$^{-}$ located near the solute, even if small in amplitude, are very effective in modulating $\Delta E$, compared with those of EMI$^+$ with extended structure and charge distribution.  This also explains why the anion dominance in short-time dynamics is reduced considerably in EMI$^+$PF$_6^-$ (although the anion contributions tend to be still larger than the corresponding cation contributions); translations of bulky PF$_6^-$ do not modulate $\Delta E$ as much as those of Cl$^{-}$ because the former are located at a larger distance from the solute than the latter.  In addition, due to the enhancement in the cation structure around the solute in EMI$^+$PF$_6^-$ compared to that in EMI$^+$Cl$^-$ [Sec.~III\,A of I], EMI$^+$ plays a more significant role in the former than in the latter. 
Another noteworthy aspect is that the relaxation of the cross correlation between the anions and cations is slow.  This is presumably due to the local electroneutrality condition that strongly correlates anions and cations.  Finally, we note that as in the solvent spectral shifts studied in I (see Table~II there), the role of anions in equilibrium solvation dynamics becomes somewhat reduced 
in the presence of benzenelike solutes (results are not shown here) compared to diatomic solutes.  We nonetheless notice that the general characteristics of $C_{a/b}(t)$ are very similar between the two solute types.  


The time-dependent friction $\zeta(t)$ for solvent dynamics in the GLE description (Sec.~\ref{sec:theory}) is exhibited in Fig.~\ref{fig:fric}.  To obtain $\zeta(t)$, we have first computed $C_{a/b}(t)$ and $\Phi(t)$ and used their Fourier-Laplace transforms in the second and third expressions of Eq.~(\ref{eq:zeta:phi}) to determine $\tilde \zeta[z]$ for small and large $z$, respectively.  These two functions have been connected smoothly, so that resulting $\tilde \zeta[z]$ when substituted into Eq.~(\ref{eq:cofz}) reproduces $C(t)$ accurately in the entire time domain.\cite{hynes:solvation,comment:friction}
The results for EMI$^+$Cl$^-$ in the presence of diatomic solutes show that analogous to $C_{a/b}(t)$, there are two totally different dynamic regimes for $\zeta(t)$, ultrafast relaxation with a time scale shorter than 0.1~ps and an extremely slow decay.   Due to numerical uncertainties involved in the determination of $\zeta(t)$, it is not easy to determine unambiguously if the latter is a single-exponential decay.  Nevertheless, with a single-exponential fit,  the characteristic time associated with the long-time tail of $\zeta(t)$ would be $\sim\! 40$\,ps for NP and $\sim\! 100$\,ps for IP.  This reveals that the memory effects on solvation dynamics last for an extended period in RTILs.  As such, the effective friction on solvent motions varies considerably with their time scale and makes the long-time behavior of associated $C_{a/b}(t)$ non-exponential.  This could have important consequences for reaction dynamics of various charge shift and transfer processes.  For example, in the case of electron transfer reactions, the effective friction relevant to barrier crossing could vary substantially with the barrier frequency in RTILs.\cite{hynes:et}

To place our results for equilibrium solvation dynamics  in perspective, we digress here to consider potential limitations imposed by the rigid, united-atom description used in the present study.  While our comparative analysis in I shows that it captures static solvation properties well, we expect that solvation dynamics---both ultrafast and long-time components--- in real ionic liquids would be accelerated compared to our predictions through participation of cation (and to some extent, PF$_6^-$) internal motions that are absent in the rigid, united-atom description.  To check this, 
we have conducted a test simulation using the flexible, all-atom description of Ref.~\onlinecite{ref:margulis} for EMI$^+$ and the rigid, all-atom description of Ref.~\onlinecite{kaminski} as in I. 
In Fig.~\ref{fig:allatom}(a), its preliminary results for $C_{a/b}(t)$, averaged over a 2~ns trajectory in the presence of the diatomic solute are compared with the results of the united atom description.  While long-time dynamics of the all-atom description with flexible cations are faster than those of the rigid, united-atom model,  the presence of clear demarcation between the subpicosecond and diffusive regimes is common to both models.  Furthermore, the relative contributions ($\sim\!60$--70~\%) and time scale ($\sim\!0.5$~ps) of the subpicosecond dynamics are essentially the same between the two.  It is quite remarkable that two models with rather different parametrization and representation yield this level of quantitative agreement in their TCF predictions. 
One of their most noticeable differences though lies in the initial relaxation near $t=0$, which is considerably faster for the flexible cation model than for the rigid one.  
For better understanding, we decompose $C_{\text{IP}/\text{NP}}(t)$ of the former into the cation and anion components in Fig.~\ref{fig:allatom}(b) just like the rigid case above.  Comparison with Fig.~\ref{fig:coft:comp}(d) shows that it is indeed the cation component of the flexible model that is responsible for its rapid oscillatory decay in the first $\sim\! 0.1$\,ps.\cite{anion:internal}  The oscillation period is $\sim\! 30$\,fs, which corresponds to a frequency of $\sim\! 1000$--1200\,cm$^{-1}$.  Its likely sources are bending motions of various CH groups of EMI$^+$. 
While the internal modes of this frequency may not be excited near equilibrium at room temperature, they will nonetheless participate in nonequilibrium relaxation with excess energy (see Figs.\ \ref{fig:motion} and~\ref{fig:deltae:ion} below).  More generally, our classical analysis here, though preliminary, shows that ion internal motions, e.g., torsion and bending modes of polar groups, could be important to energetics and dynamics in RTILs.  Thus while the overall aspects of $C_{a/b}(t)$ can be captured quite reasonably by rigid potential models, a proper account of ion internal motions will be needed to accurately quantify solvation dynamics and related dynamic electronic spectroscopies, e.g., optical Kerr effect spectroscopy.\cite{kerr:exp}

\subsection{Nonequilibrium Solvation Dynamics}
\label{subsec:noneq}

The MD results for $S_{a/b}(t)$ given by Eqs.\ (\ref{eq:stokes:response}) and~(\ref{eq:stokes:gap}) are shown in Fig.~\ref{fig:soft}.  Just like $C_{a/b}(t)$, $S_{a/b}(t)$ is characterized by at least two distinct relaxation processes in all cases we studied.  About 50--70\,\% of the entire solvent relaxation occurs in less than 0.3\,ps, irrespective of the probe solute.  This confirms the earlier conjecture based on dynamic and static Stokes shift measurements\cite{ref3,ref4} that a significant portion of solvent relaxation occurs on a subpicosecond time scale in RTILs.   We notice in Fig.~\ref{fig:soft} that subpicosecond dynamics of $S_{\text{NP}/\text{IP}}(t)$ (and $S_{\text{RB}/\text{DB}}(t)$) are faster than those of  $S_{\text{IP}/\text{NP}}(t)$ (and $S_{\text{DB}/\text{RB}}(t)$).  We will return to this point later on.  

In Figs.\ \ref{fig:soft:comp:emicl} and~\ref{fig:soft:comp:emipf6}, $S_{a/b}^{\text{an}}(t)$ and $S_{a/b}^{\text{cat}}(t)$, the anion and cation components of $S_{a/b}(t)$, are presented for EMI$^+$Cl$^-$ and EMI$^+$PF$_6^-$.   While both the anions and cations play an important role in relaxation dynamics, the contributions from the former are much larger than those from the latter in EMI$^+$Cl$^-$.   As mentioned above, this is due to the small size of $\mathrm{Cl^-}$.  For EMI$^+$PF$_6^-$ with bulky anions, the contributions from the anions and cations are comparable.  We nonetheless notice that the anion contributions to the subpicosecond solvent dynamics, measured as the initial drop in $S_{a/b}^{\text{an}}(t)$ in the first 0.2\,ps or so, tend to be somewhat larger than the cation contributions.
Even in the case of the benzenelike solutes in Figs.\ \ref{fig:soft:comp:emipf6}(c) and~\ref{fig:soft:comp:emipf6}(d) where $S_{a/b}^{\text{cat}}(0)$ is larger than $S_{a/b}^{\text{an}}(0)$, the anion contributions to the subpicosecond dynamics are still larger than the corresponding cation contributions.  Another noticeable feature is that as in the $C_{a/b}(t)$ case,  the relative contributions of cations and anions to $S_{a/b}(t)$ vary with the solute type; the cations tend to play a more important role in the presence of the benzenelike solute than in the presence of the diatomic solute (see above).  In I, it is conjectured that
variations in solute structure and charge distributions influence respective roles of cations and anions and thus details of solvation by modulating the ion distributions close to the solute.
This generally supports the view that solvation properties of ionic liquids measured via dynamic and static electronic spectroscopies can vary with the probe solutes.\cite{petrich}  Nevertheless, overall features of $S_{a/b}(t)$ in Fig.~\ref{fig:soft} do not seem to be that sensitive to the solute types although our investigation is limited to the diatomic and benzenelike solutes. 

Before we proceed further, we briefly pause here to consider MD statistics employed for nonequilibrium calculations.  We note that at $t=0$, 
the anion and cation components of $S_{a/b}(t)$ satisfy
\begin{equation}
S_{a/b}^{\text{cat,an}}(0) = S_{b/a}^{\text{cat,an}}(0)\ .
\label{eq:component:relation}
\end{equation}
This is due to the fact that despite the nonequilibrium character of $S_{a/b}(t)$ and its components, their initial values depend only on equilibrium quantities [cf.\ Eqs.\ (\ref{eq:stokes:response}) and~(\ref{eq:ave:relation})].  
We notice in Figs.\ \ref{fig:soft:comp:emicl} and~\ref{fig:soft:comp:emipf6} that Eq.~(\ref{eq:component:relation}) is well  satisfied by our MD results.  This indicates that 
400 configurations used to compute averages provide decent statistics for nonequilibrium.

Returning to our main thread, we decompose cation motions into translation and rotation and study their respective contributions to the solvent relaxation.\cite{steele,ladanyi:motion,schwartz}  For this purpose, it is convenient to consider 
$\overline{\Delta\Delta E_{a\to b}}(t) \equiv \overline{\Delta E_{a\to b}}(t) - \overline{\Delta E_{a\to b}}(0)$, which starts from the null value at $t=0$, for each type of motion.  According to the results in Fig.~\ref{fig:motion}, the contributions from cation translations exceeds those from cation rotations by a factor of two or more, revealing that translational motions of ions are of primary importance in $S_{a/b}^{\text{cat}}(t)$ throughout the entire relaxation.\cite{comment:flex}  This holds for all cases we have studied. 

In view of huge electrostrictive effects found in I, it is needless to say that the nonequilibrium solvent relaxation is accompanied by significant solvent rearrangements in RTILs.  To understand the link between the structural relaxation and solvation dynamics, we examine the time-evolution of solvent structure subsequent to an instantaneous change in the solute charge distribution at $t=0$.  
The results for the evolving radial distributions of the cations and anions around the diatomic solute in $\mathrm{EMI^+Cl^-}$ are exhibited in Fig.~\ref{fig:gofrt}.  
We first consider Figs.\ \ref{fig:gofrt}(a) and~\ref{fig:gofrt}(b), where the solvent, initially equilibrated to the IP solute, relaxes to a new state in equilibrium with the NP charge distribution.  This corresponds to $S_{\text{NP}/\text{IP}}(t)$ since the solvent relaxation occurring in the presence of the active NP state is monitored via the $\mathrm{NP\to IP}$ electronic transitions.  The time evolution of the number of Cl$^-$ in the first solvation shell, viz., within 5\,\AA\ of the positive site of the solute is given in Table~\ref{table:cl}.  
For convenience, we will refer to this quantity as the Cl$^-$ coordination number hereafter.  We notice that the structural change in the first $\sim 0.2$\,ps is not that significant on a relative scale.  
During this period, the evanescent first solvation shell moves out a little and becomes broadened somewhat; its peak height decreases by $\sim\! 35$\%.  The reduction in the Cl$^-$ coordination number is less than unity from 4.1 to 3.4.  
Only after $\sim\!1$\,ps, does the destruction of the first solvation shell become obvious.  
Despite this, $S_{\text{NP}/\text{IP}}(t)$ relaxes by more than 60\,\% during the first $\sim 0.2$\,ps as shown in Figs.\ \ref{fig:soft} and~\ref{fig:soft:comp:emicl}.  In terms of energy, this corresponds to a change of $\sim\! 30\, \mathrm{kcal/mol}$ in $\Delta E_{\text{NP}\to\text{IP}}^{\text{an}}$  [cf.\ Fig.~\ref{fig:motion}(a)].  

To gain further insight into the ultrafast relaxation of $S_{a/b}(t)$, we divide anions into the first solvation shell and ``bulk'' region\cite{comment:bulk} at $t=0$ and investigate the temporal behavior of their respective contributions to $\Delta E_{a\to b}$.   The MD results corresponding to $S_{\text{NP}/\text{IP}}(t)$ are given in Fig.~\ref{fig:deltae:ion}(a), where the cation results are also presented for comparison.  The contributions from the first solvation shell anions vary significantly with time, especially at small $t$, whereas those from the bulk change little.  This reveals that the contributions to subpicosecond dynamics of $S_{\text{NP}/\text{IP}}^{\text{an}}(t)$ arise almost completely from Cl$^-$ in the first solvation shell!  Though lesser in extent, this is also the case with cations.  This paints a picture that motions of a few ions close to the solute essentially govern the fast nonequilibrium solvation dynamics occurring immediately after the instantaneous $\mathrm {IP\to NP}$ change in the solute charge distribution.  It is remarkable that these few ions account for energy relaxation of $\sim\! 40\,\mathrm{kcal/mol}$ in about $0.2$\,ps, which corresponds to more than 60\,\% of the entire $S_{\text{NP}/\text{IP}}(t)$ decay.  

We investigate the time evolution of the average distance between the solute and ions, which are initially located in the first solvation shells.  The results in Fig.~\ref{fig:distance}(a) show that ion translations are characterized by two different regimes: The displacement at short time grows more rapidly than a linear increase in $t$.    We ascribe this to ``inertial'' translations initiated by sudden changes in the Coulomb force on ions exerted by the solute at $t=0$.  The displacement at later time increases nearly linearly with $t$, suggesting a diffusive transport regime for ions, subject to the Coulomb force (thus biased).  It should be noticed that the average displacement of the first solvation shell ions in the inertial regime is not significant; their net displacement during the first 0.2\,ps is about 0.5\,\AA.  But due to their strong Coulombic interactions with the solute, this is sufficient to modulate and relax $\Delta E_{\text{NP}/\text{IP}}$ by more than 60\,\%.     
We thus conclude that in the case of the initial $\mathrm {IP \to NP}$ change in the solute electronic state and ensuing solvent relaxation, rapid inertial translations of small amplitude  associated with the ions situated very close to the solute are responsible for the ultrafast subpicosecond dynamics of $S_{\text{NP}/\text{IP}}(t)$.\cite{comment:flex}  The subsequent slow decay of $S_{\text{NP}/\text{IP}}(t)$ is attributed to (biased) diffusive migrations of these ions into the ``bulk'' region\cite{comment:bulk} as well as diffusive rearrangements of bulk ions. 

We now turn to the opposite case, where the solvent, initially equilibrated to the NP solute, readjusts to the the newly-created IP charge distribution.  Analogous to the $S_{\text{NP}/\text{IP}}(t)$ case analyzed above, the initial relaxation of the Stokes shift function $S_{\text{IP}/\text{NP}}(t)$ is extremely fast.  The $t$-dependent solvent distributions in Figs.\ \ref{fig:gofrt}(c) and~\ref{fig:gofrt}(d), on the other hand, show that the formation of solvent structures around the solute is quite sluggish compared to the short-time relaxation of $S_{\text{IP}/\text{NP}}(t)$.  To be specific, the nascent first solvation shells develop a descernible structure only after $\sim\! 0.4$\,ps.   During this period, the Cl$^{-}$ coordination number changes just a little  from 1.8 to 2.15 (see Table~\ref{table:cl}).  In the first $\sim\! 0.2$\,ps, during which more than 50\,\% of the $S_{\text{IP}/\text{NP}}(t)$ relaxation is completed [Fig.~\ref{fig:soft}(a)], the increase in the Cl$^{-}$ coordination number is mere 0.2!  For comparison, recall that the Cl$^{-}$ coordination number change during the same 0.2\,ps period is $\sim\!0.7$  in the case of $S_{\text{NP}/\text{IP}}(t)$.  Thus while rapid inertial translations of ions (see Fig.~\ref{fig:distance}(b)) are expected to be responsible for the subpicosecond relaxation of $S_{\text{IP}/\text{NP}}(t)$ as in $S_{\text{NP}/\text{IP}}(t)$, the ions close to the solute alone may not be sufficient to account for the ultrafast energy relaxation of 30--40\,kcal/mol.

To pursue the last point above a little further, we group the anions together in three different regions according to their separation from the solute at $t=0$ and analyze their contributions to $\Delta E_{a\to b}(t)$ in Fig.~\ref{fig:deltae:ion}(b).  For convenience, we refer to the three regions as ``near'' ($r<5$\,\AA),\cite{comment:near} ``intermediate'' ($5\,\mathrm{\AA} <r<7$\,\AA), and ``far''  ($r>7$\,\AA) regions, where $r$ is the initial distance between the solute positive site and Cl$^{-}$.  For completeness, the cation results are also shown there.\cite{comment:cation:region}
We notice that anions from all three regions make substantial contributions to subpicosecond solvation dynamics.  For example, the contribution from Cl$^{-}$ in the far region exceeds 10\,\% of the total anion contribution to ultrafast $S_{\text{IP}/\text{NP}}(t)$.  This is in contrast with the $S_{\text{NP}/\text{IP}}(t)$ case above, where short-time dynamics are almost completely governed by ion motions in the near region [Fig.~\ref{fig:deltae:ion}(a)].   This is due to the difference in the anion density in the near region between the two cases (cf.\ Fig.~\ref{fig:gofrt}).  In the $S_{\text{NP}/\text{IP}}(t)$ case, the number of Cl$^{-}$ initially in the near region is high, so that the large magnitude of their Coulombic interactions with the solute dominates $\Delta E_{\text{NP}\to\text{IP}}$.   By contrast, in the case of $S_{\text{IP}/\text{NP}}(t)$,  due to the absence of Coulombic interactions between the NP solute and ions, there are on the average fewer than two Cl$^{-}$ ions present near the solute at the outset of the solvent relaxation.  These are not sufficient to dissipate the extra Coulomb energy generated by the change in the solute charge distribution at $t=0$.  Therefore the Cl$^{-}$ in the intermediate and far regions play a considerably more signficant role in the subpicosecond relaxation of $S_{\text{IP}/\text{NP}}(t)$ than in the case of $S_{\text{NP}/\text{IP}}(t)$.  This also explains why the former are somewhat slower than the latter as we noticed in Fig.~\ref{fig:soft} above.  Since the electrostatic force on ions arising from the solute decreases rapidly with their separation, it will take longer for ions at a large distance from the solute to dissipate energy through translations than those close to the solute.  Slow relaxation of $S_{\text{IP}/\text{NP}}^{\text{an}}(t)$ occuring on a time scale longer than $1$~ps mainly involves migration and redistribution of Cl$^{-}$ originating from the intermediate and far regions.   

To summarize, while the basic features of the Stokes shift functions $S_{\text{NP}/\text{IP}}(t)$ and $S_{\text{IP}/\text{NP}}(t)$ are very similar, the range of Coulombic interactions that determine the energy gap relaxation differs considerably between the two.  To be specific, in the case of $S_{\text{NP}/\text{IP}}(t)$, rapid receding of the first solvation shell ions from the solute essentially determines the subpicosecond solvation dynamics; thus the solute-solvent Coulombic interactions at close ranges are of primary importance.  By contrast, in the  $S_{\text{IP}/\text{NP}}(t)$ case, Fig.~\ref{fig:deltae:ion}(b) shows that ions at all distances contribute to the solvation dynamics  through their initial approach to the solute (``clustering in'') via small amplitude inertial translations.  Therefore the Coulombic interactions between the solute and solvent relevant for $S_{\text{IP}/\text{NP}}(t)$ dynamics are of considerably longer range than those for $S_{\text{NP}/\text{IP}}(t)$.  (For an analysis of related issues in dipolar liquids, see Ref.~\onlinecite{schwartz}.)
While our conclusion is based exclusively on the analysis of EMI$^+$Cl$^-$, it also applies to EMI$^+$PF$_6^-$.  The major difference between the two is the relative importance of cations; i.e., they play a more significant role in the latter than in the former.  

Before we conclude, we consider the validity of linear response in RTILs.  Comparison between $C_{a/b}(t)$ and $S_{a/b}(t)$ in Fig.~\ref{fig:comp} shows that despite the large electrostrictive effects observed in I (Figs.\ 2--6 there) and the difference in the range of Coulombic interactions relevant to $S_{a/b}(t)$ found above, linear response holds surprisingly well  in EMI$^+$Cl$^-$.  In particular, $C_{\text{IP/NP}}(t)$ and $S_{\text{NP/IP}}(t)$ yield an excellent agreement in short-time dynamics and similarly for other pairs, $C_{\text{DB/RB}}(t)$ and $S_{\text{RB/DB}}(t)$, etc.  This is presumably due to similarity in the solvent configurations ${\cal Q}_t$ between $C_{a/b}(t)$ and $S_{b/a}(t)$ for small $t$ [cf.\ Eqs.\ \ref{eq:eq:av} and~\ref{eq:stokes:gap}], so that their early dynamics occur in a similar solvent environment. (Recall that the two are characterized by the same ensemble distribution function, i.e., their solvent configurations are the same at $t=0$.)  In a similar vein, we would expect that at long time, slow dynamics of $S_{b/a}(t)$ would resemble those of $C_{b/a}(t)$ (rather than those of $C_{a/b}(t)$) because ${\cal Q}_t$ associated with $S_{b/a}(t)$ eventually evolves to the configurations equilibrated with the solute $b$-state (see Fig.~\ref{fig:gofrt} above).  While this appears to be the case in Fig.~\ref{fig:comp}, we need to perform nonequilibrium simulations considerably longer than 2~ps to test this notion unambiguously.  Finally, we note that the electronic polarizability absent in the current study---especially that of the probe solutes---will introduce a significant nonlinear behavior in solvent response.\cite{kim:md}

\section{Concluding Remarks}

In the present article, we have studied dynamic properties of solvation in EMI$^+$Cl$^-$
and EMI$^+$PF$_6^-$ via MD simulations.  We have investigated dynamics Stokes shifts subsequent to an instantaneous change in the solute charge distribution and accompanying solvent structure reorganization that evolves in time.  We have studied fluctuating solvent dynamics near equilibrium and analyzed associated friction dynamics via the generalized Langevin equation.  It was found that both equilibrium and nonequilibrium 
dynamics show two totally different relaxation processes: rapid subpicosecond dynamics arising mainly from inertial ion translations and slow relaxation ascribed to ion transport.
While the solvent dynamics are dominated by Cl$^-$ motions in EMI$^+$Cl$^-$, both anions and cations play an important role in EMI$^+$PF$_6^-$.  We have also found that linear response holds reasonably well in both ionic liquids.  

We have examined molecular details of nonequilibrium solvent relaxation by analyzing cation and anion motions and their contributions from different solvation regions.  We found that cation translations are considerably more important than cation rotations during the entire solvent relaxation.\cite{comment:flex} 
The range of solvation regions which participate in the subpicosecond relaxation varies significantly with the initial solvent density near the solute.  In the case of a high local density near the solute at $t=0$, ensuing subpicosecond solvent relaxation is mainly governed by the motions of a few ions close to the solute.  By contrast, in the opposite case of a low initial density, solvent ions not only close to but also relatively far from the solute contribute to the subpicosecond relaxation dynamics.  Despite this difference, their dynamic Stokes shift functions show very similar characteristics.

It would be worthwhile in the future to extend the present study based on a rigid, united-atom description in more realistic directions.  Among others, the incorporation of molecular flexibility into simulations in the all-atom representaion would be desirable.  Our preliminary results considered here indicate that though overall dynamic features are well captured by the rigid model, it clearly misses several important details, in particular, at short time, due to the absence of ion internal motions, such as torsion and bending.   
Another aspect not included in our model description is electronic polarizability of ions.  
While its effect in RTILs may not be as significant as in polar solvents, its proper account is needed to accurately quantify e.g., solvent spectral shifts and reorganization free energy.   

Finally, solute dynamic properties, such as rotational friction and vibration energy relaxation, in RTILs and their variations with solute electronic structure will be reported elsewhere.\cite{rtil:rot:vib}

\acknowledgments
This work was supported in part by the Ministry of Education of Korea through the BK21 Program, 
by KOSEF Grant No.\ 01-2002-000-00285-0, and by NSF Grant No.\ CHE-0098062.


\begin{thebibliography}{<100>}
\bibitem{comment:solute:parameter} 
These parametric values for the diatomic solute are exactly the same as in Ref.~\onlinecite{shim}.
\bibitem{comment:dh}
The Debye-H{\"u}ckel theory, which is valid only in the infinite dilution limit of ionic solutions, predicts monotonically decreasing $Q_\alpha(r)$
\begin{equation}
Q_\alpha(r)= \frac{A_\alpha}{r} \mathrm{e}^{- r/\lambda_{\text{DH}}} \ ;\quad
\lambda_{\text{DH}}=\frac{k_{\text{B}} T}{4\pi \sum_\alpha n_\alpha q_\alpha^2}\ .
\end{equation} 
Its direct application to the RTILs would yield
$\lambda_{\text{DH}}\approx ??$~\AA.
\bibitem{comment:force:trend}
This was first noted in Ref.~\onlinecite{carter:hynes}.
\bibitem{comment:force:trend2}
 The opposite $k_{a/b}$ trend in water  can be understood as follows.
The effective size of hydrogen in water molecules is small; i.e., the molecular shape of water based on its LJ parameters is quite spherical.  Thus as the local solvent density increases due to electrostriction, fluctuations in orientations of water molecules near the solute via librational motions are not as sterically hindered as non-spherical molecules such as acetonitrile.  Furthermore, the electrostatic interaction strength variation in water would be larger due to (partial) hydrogen bonding formation with dipolar solutes than that in aprotic solvents.  Therefore the electrostatic interaction strength is a dominant factor for  $\langle(\delta\Delta E_{a\to b})^2\rangle$ in the case of water.
\bibitem{comment:noneq:relax}
Thus for $S_{a/b}(t)$, the time-evolution of ${\cal Q}_t$ occurs in the presence of the solute $a$-state, while the initial ${\cal Q}_0$ is equilibrated to state $b$.
\bibitem{comment:accuracy}
Within the numerical accuracy of 1~kcal/mol which is the bin size employed for $\Delta E_{a\to b}$ to calculate $P_{a/b}$ and $F_{a/b}$ (Eqs.\ (\ref{eq:probability}) and~(\ref{eq:free:energy})), our MD results for the $\Delta E_{a\to b}^{\text{min}}$ values and $\langle\Delta E_{a\to b}\rangle$ are the same.
\bibitem{water}
In view of the discussions above, it would be the opposite in the case of water, e.g.,
$\lambda_{\text{IP}/\text{NP}}$ is smaller than $\lambda_{\text{NP}/\text{IP}}$.
\bibitem{comment:solvation:shell}
The first solvation shells for anions and cations (and their corresponding bulk regions) are defined with respect to the positive and negative site of the dipolar solute, respectively.
\bibitem{comment:firstshell}
These ions mainly correspond to the anions and cations of the first solvation shells, which are defined with respect to the positive and negative sites of the solute.
\bibitem{comment:cation:rot}
By decomposing the cation motions into rotational and translational components, we have investigated their respective contributions to $\Delta E$ relaxation.  We have found that the contribution from rotational motions is considerably larger than that from translations.
\bibitem{comment:flex}
However, there is a caveat:  the internal motions of anions and cations, e.g., bending and torsional motions of charge and polar groups of EMI$^+$, absent in the current rigid model description can play an important role, especially in very early dynamics [cf.\ Fig.~\ref{fig:allatom}].  



the internal motions of cations are totally absent in the rigid model description we employed.  In Sec.~\ref{subsec:eq} above, we observed that the initial relaxation of $C_{a/b}(t)$ [Fig.~\ref{fig:allatom}] of the flexible model during the first $\sim\!50~fs$ is considerably fast compared to that of the rigid model and attributed it 
to the internal motions, in particular, bending and torsional motions of charge and polar groups.  By analogy, we believe that $S_{a/b}^{\text{cat}}(t)$ and thus $S_{a/b}(t)$ should have a similar ultrafast component, which is not captured in the current rigid description.
\bibitem{keblinski}
Similar results were obtained also in the MD simulations of NaCl.
See Ref.~\onlinecite{molten_salt:sim}.
\bibitem{comment:near}
This region is referred to as the first solvation shell above.
\bibitem{comment:cation:region}
The near, intermediate, and far regions for the cations are defined as $r<6$\,\AA, $6\,\mathrm{\AA} <r<8$\,\AA\ and  $r>8$\,\AA, respectively, where $r$ is the distance between the cation center and solute (+) site.
\bibitem{comment:bulk}
The bulk is defined to be the rest of the anions (cations) not included in the first solvation shell.
\bibitem{anion:internal}
Though neglected here, anion internal motions, i.e., PF stretch and bend, would introduce similar rapid oscillations (of $\sim\! 40$--60\,fs) to the anion component and thus to $C_{\text{IP}/\text{NP}}(t)$. We nonetheless believe that their contribution would be smaller than that of cation internal motions because for a given frequency, the average amplitude of internal motions associated with more massive F is considerably smaller than that associated with lighter H.
\end{thebibliography}

\newpage

\begin{table*}
\centering
\caption{\footnotesize{MD results$^{a)}$}}
\vspace{10pt}
\begin{tabular}{|cc|c|c|c|c|c|c|}
\hline
solvent\ &\  density\ &\ $a/b$\ &\ $\langle (\delta \Delta E_{a\to b})^{2} \rangle$\ \ &\ $\langle (\delta \Delta \dot{E}_{a\to b})^{2} \rangle$\ \ &\ $\omega_s$\,(ps$^{-1}$)\ &\ $\beta$$^{\,b)}$\ &\ $\tau_0$$^{\,b)}$\\
\hline
\ EMI$^+$Cl$^-$\ &\ 1.1\ &\ NP/IP\ &\ 83.3\ &\ 7520\ &\ 9.5\ &\ 0.25\ &\ 0.31\\
  &\ &\ IP/NP\ &\ 60.5\ &\ 14720&\ 15.6\ &\ 0.11\ &\ 0.15\\
\ EMI$^+$Cl$^-$\ &\ 1.1&\ RB/DB\ &\ 77.5\ &\ 7140\ &\ 9.6\ &\ 0.5\ &\ 0.76\\
  &\ &\ DB/RB\ &\ 54.2\ &\ 13530\ &\ 15.8\ &\ 0.1\ &\ 0.49\\
\ EMI$^+$Cl$^-$\ &\ 1.2\ &\ NP/IP\ &\ 72.3\ &\ 10590\ &\ 12.1\ &\ 0.12\ &\ 0.14\\
  &\ &\ IP/NP\ &\ 57.9\ &\ 15010&\ 16.1\ &\ 0.05\ &\ 0.005\\
\ EMI$^+$PF$_6^-$\ &\ 1.31\ &\ NP/IP\ &\ 79.8\ &\ 2870\ &\ 6.0\ &\ 0.25\ &\ 0.69\\
 &\ &\ IP/NP\ &\ 64.5\ &\ 3820\ &\ 7.7\ &\ 0.15\ &\ 0.13\\
\ EMI$^+$PF$_6^-$\ &\ 1.375\ &\ NP/IP\ &\ 77.8\ &\ 3290\ &\ 6.5\ &\ 0.2\ &\ 0.54\\
 &\ &\ IP/NP\ &\ 52.0\ &\ 4310\ &\ 9.1\ &\ 0.1\ &\ 0.077\\
\ EMI$^+$PF$_6^-$\ &\ 1.375\ &\ RB/DB\ &\ 81.2\ &\ 4810\ &\ 7.7\ &\ 0.6\ &\ 2.22\\
 &\ &\ DB/RB\ &\ 38.7\ &\ 5120\ &\ 11.5\ &\ 0.05\ &\ 0.0935\\
\ CH$_3$CN$^{\,c)}$\ &\ 0.73\ &\ NP/IP\ &\ 46.4\ &\ 4640\ &\ 10.0\ && \\
 &\ &\ IP/NP\ &\ 41.3\ &\ 6870\ &\ 12.9\ && \\
\hline
\end{tabular}
\label{table:result}
\end{table*}

\vspace*{-20pt}\noindent
$^{a)}$Units for density, $\Delta E_{a\rightarrow b}$ and time ($t$ and $\tau_0$) are g/cm$^{3}$, kcal/mol and ps, respectively.

\noindent
$^{\,b)}$A stretched exponential fit to $C_{a/b}(t)$, i.e., $\exp[-(t/\tau_0)^\beta]$, for $t\le100$\,ps after ultrafast initial relaxation.

\noindent
$^{c)}$MD results at $T=300$\,K.  The LJ parameters and partial charges employed are from Ref.~\onlinecite{acetonitrile}.

\newpage

\begin{table}
\caption{Time evolution of the Cl$^-$ coordination number around the diatomic solute after the change in the solute charge distribution}
\vspace*{10pt}
\begin{tabular}{|cc|c|c|c|c|c|c|c|c|c|}
\hline
&\ $t$\,(ps)\ &\ 0.0\ &\ 0.1\ &\ 0.2\ &\ 0.4\ &\ 0.6\ &\ 0.8\ &\ 1.0\ &\ 2.0\ &\ $\infty$ \\
\noalign{\vspace*{-5pt}}
\ active state& site &&&&&&&&&\\
\hline
\ NP &\ (+)\ &\ 4.07\ &\ 3.41\ &\ 3.40\ &\ 3.35\ &\ 3.28\ &\ 3.19\ &\ 3.04\ &\ 2.55\ &\ 1.80 \\
 &\ $(-)$\ &\ 0.34\ &\ 0.43\ &\ 0.71\ &\ 1.25\ &\ 1.58\ &\ 1.54\ &\ 1.71\ &\ 1.79\ &\ 1.80 \\
\ IP &\ (+)\ &\ 1.80\ &\ 1.77\ &\ 2.00\ &\ 2.15\ &\ 2.31\ &\ 2.39\ &\ 2.46\ &\ 2.87\ &\ 4.07 \\
 &\ $(-)$\ &\ 1.80\ &\ 1.45\ &\ 1.22\ &\ 1.02\ &\ 0.88\ &\ 0.86\ &\ 0.86\ &\ 0.65\ &\ 0.34 \\
\hline
\end{tabular}
\label{table:cl}
\vspace*{4in}
\end{table}

\newpage
\begin{figure}
\caption{Equilibrium time correlation function $C_{a/b}(t)$ in (a) EMI$^+$Cl$^-$ at density $\rho=1.1$\,g/cm$^{3}$ and in (b)~EMI$^+$PF$_6^-$ at $\rho=1.375$\,g/cm$^{3}$ for the solute active/reference states $a/b$: NP/IP (---); IP/NP $(--)$; RB/DB $(\cdots)$; DB/RB $(-\cdot-)$.}
\label{fig:coft}
\caption{Components of $C_{a/b}(t)$ for (a) NP/IP and (b) IP/NP active/reference states of the diatomic solute in EMI$^+$Cl$^-$ and for (c) NP/IP and (d) IP/NP in EMI$^+$PF$_6^-$.
The dotted and dashed lines represent the contributions from the cation and anion autocorrelations, respectively, while the dashed-dot line denotes the negative of their cross correlation divided by 2.
The RTIL densities are the same as those in Fig.~\ref{fig:coft}. }
\label{fig:coft:comp}
\caption{Time-dependent solvent friction $\zeta(t)$ (in units of ps$^{-2}$) associated with $C_{a/b}(t)$ in EMI$^+$Cl$^-$ at $\rho=1.1$\,g/cm$^{3}$: $a/b=\mathrm{NP/IP}$\,(---) and IP/NP\,$(\cdots)$.  The corresponding initial values $\zeta(0)$ are marked with $\times$ and $\square$, respectively.  Details of the initial behavior are shown in the inset.}
\label{fig:fric}
\caption{(a) Comparison of $C_{a/b}(t)$ between the united-atom (UA) and all-atom (AA) descriptions for EMI$^+$PF$_6^-$.  The solvent densities are $\rho = 1.375$\,g/cm$^{3}$ (UA) and $1.34$\,g/cm$^{3}$ (AA).  (b)~Components of $C_{\text{IP/NP}}(t)$ in the all-atom representation in (a).}
\label{fig:allatom}
\caption{Stokes shift function $S_{a/b}(t)$ in (a) EMI$^+$Cl$^-$ and (b) EMI$^+$PF$_6^-$.  The solute active/reference states employed are $\mathrm{NP/IP}$ (---), IP/NP $(--)$, RB/DB $(\cdots)$ and DB/RB $(-\cdot-)$.  The densities are the same as those in Fig.~\ref{fig:coft}.}
\label{fig:soft}
\caption{Components of $S_{a/b}(t)$ in EMI$^+$Cl$^-$ for the solute active/reference states $a/b$: (a)~NP/IP; (b)~IP/NP; (c)~RB/DB; (d)~DB/RB.
}
\label{fig:soft:comp:emicl}
\caption{Components of $S_{a/b}(t)$ in EMI$^+$PF$_6^-$ for the solute active/reference states  $a/b$: (a)~NP/IP; (b)~IP/NP; (c)~RB/DB; (d)~DB/RB.
}
\label{fig:soft:comp:emipf6}
\caption{Contributions from rotational and translational motions of ions to the relaxation of $\Delta E_{a\to b}(t)$ (in units of kcal/mol) after the initial (a) $\mathrm{IP}\!\to\! \mathrm{NP}$ and (b) $\mathrm{NP}\!\to\! \mathrm{IP}$ changes in the solute charge distribution in EMI$^+$PF$_6^-$.  The corresponding dynamic Stokes shift functions for (a) and (b) are $S_{\text{NP}/\text{IP}}(t)$ and $S_{\text{IP}/\text{NP}}(t)$, respectively.
}
\label{fig:motion}
\end{figure}
\begin{figure}
\caption{Time evolution of ion radial distributions subsequent to an instantaneous change in the solute charge distribution in EMI$^+$Cl$^-$.  The anion and cation distributions given, respectively, in (a) and (b) describe the structural relaxation accompanying $S_{\text{NP}/\text{IP}}(t)$, while the corresponding distributions in the opposite  $S_{\text{IP}/\text{NP}}(t)$ case are shown  in (c) and (d). 
For clarity, plots of the radial distribution at later time are shifted upwards from the initial distribution according to its time evolution in each pane and
only the results for the methyl group of EMI$^+$ (i.e., M1 in Fig.~1 of I) are presented in (b) and (d). }
\label{fig:gofrt}
\caption{Contributions to $\Delta E_{a\rightarrow b}(t)$ (in units of kcal/mol) from different solvent regions for (a)~NP/IP and (b) IP/NP  active/reference states of the solute.  The solid and dotted lines represent the anion and cation contributions, respectively.
In (a), EMI$^+$ ions located within 6~\AA\ from the negative site of the solute are defined as the first solvation shell, while the 5~\AA\ separation is employed to define the corresponding first solvation shell for Cl$^{-}$ around the solute positive site. }
\label{fig:deltae:ion}
\caption{Time evolution of the average distance $r$ (in units of \AA) of the ions from the solute after the instantaneous (a) $\mathrm{IP}\!\to\! \mathrm{NP}$ and (b) $\mathrm{NP}\!\to \!\mathrm{IP}$ changes in the solute charge distribution.  Only the ions located initially in the first solvation shells are considered.}
\label{fig:distance}
\caption{Comparison of equilibrium $C_{a/b}(t)$ and nonequilibrium $S_{a/b}(t)$ in EMI$^+$Cl$^-$ in the presence of (a) diatomic and (b) benzenelike solutes.  }
\label{fig:comp}
\end{figure}

\newpage

\centering 
\includegraphics[width=4.0in]{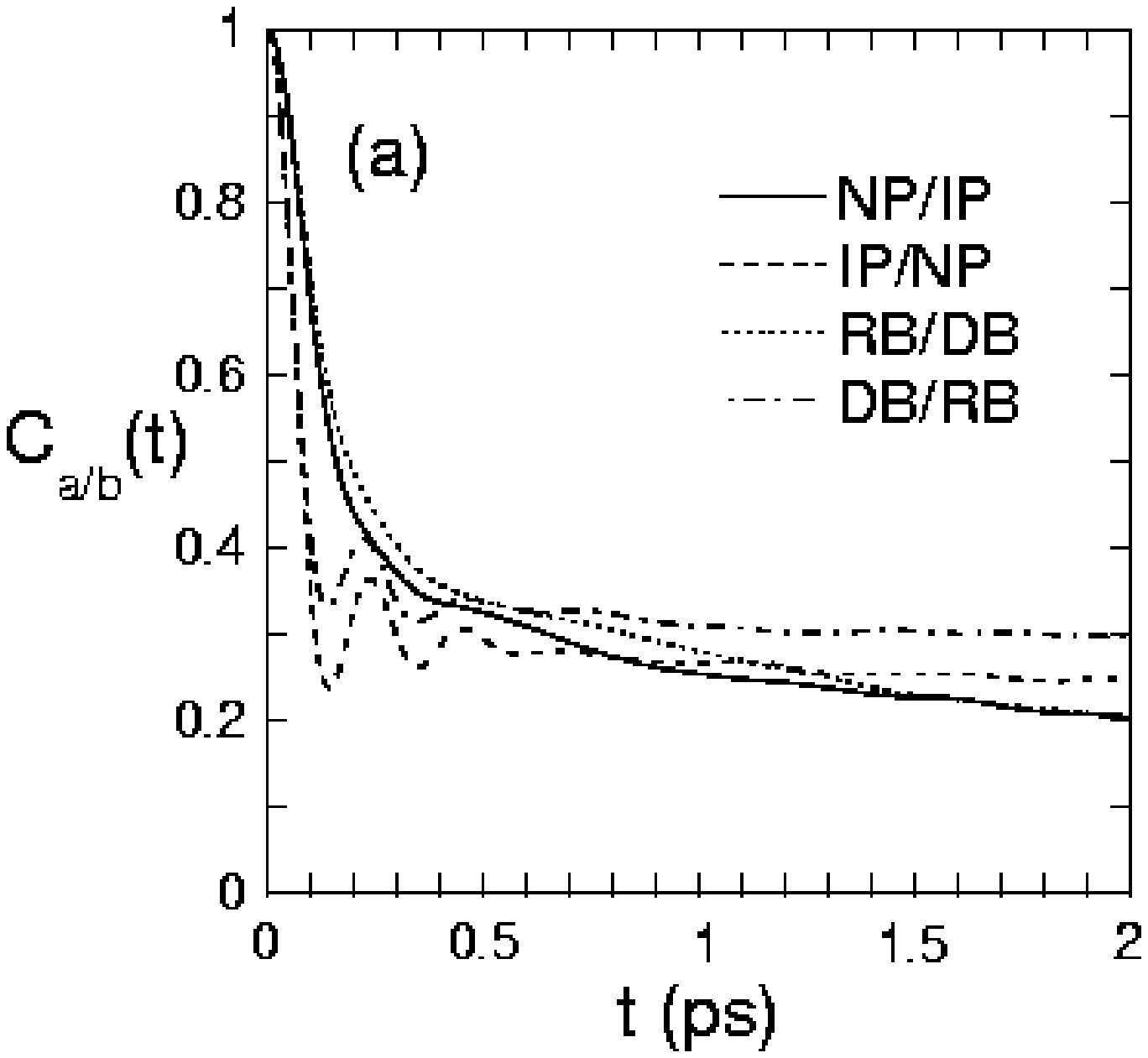}\vspace{0.3in}
\centering 
\includegraphics[width=4.0in]{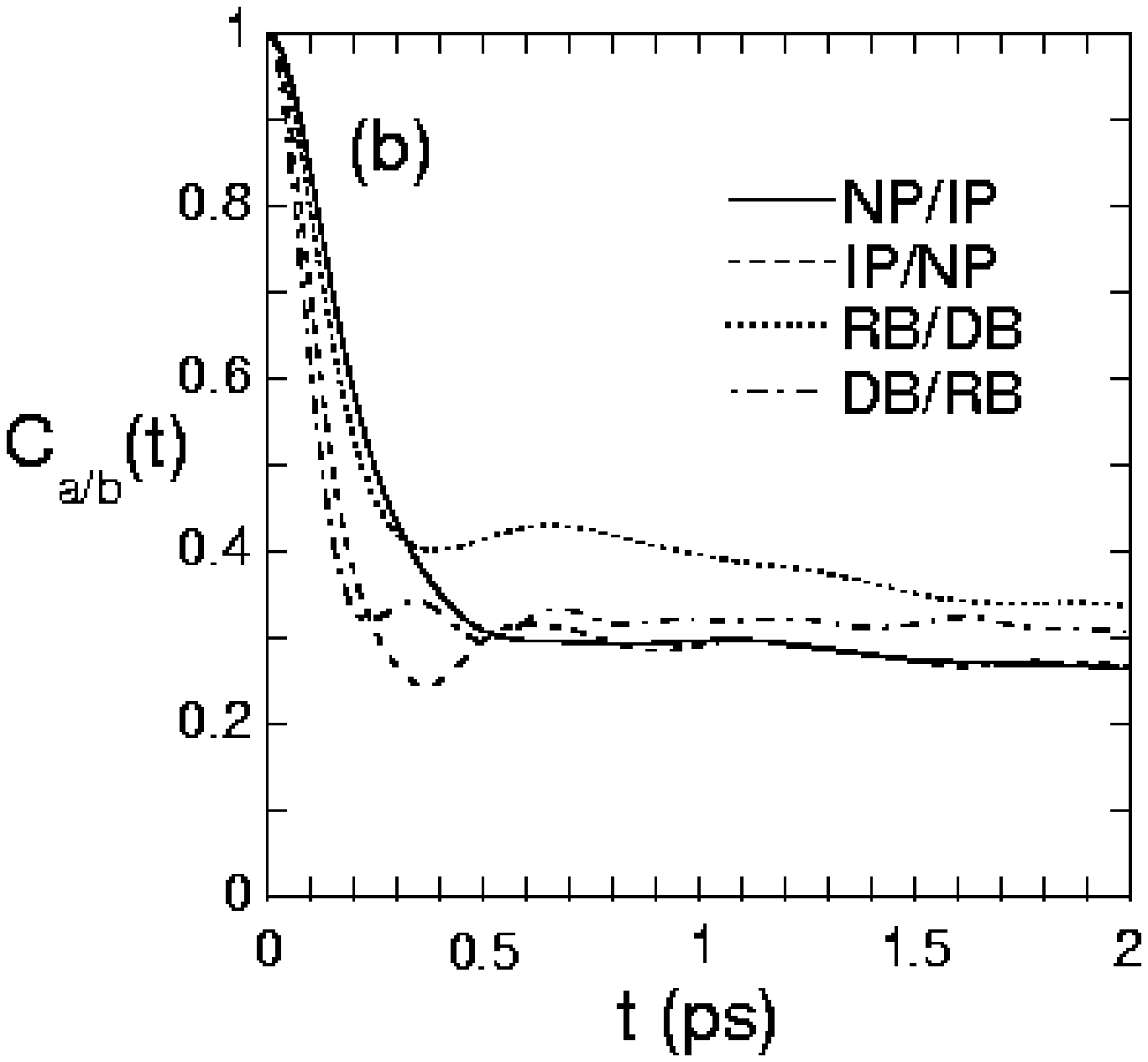}\vspace{0.3in}
\vspace{10cm}
\centerline{Fig.~\ref{fig:coft}}

\newpage

\begin{minipage}{8.0cm} 
\centering 
\includegraphics[width=3.5in]{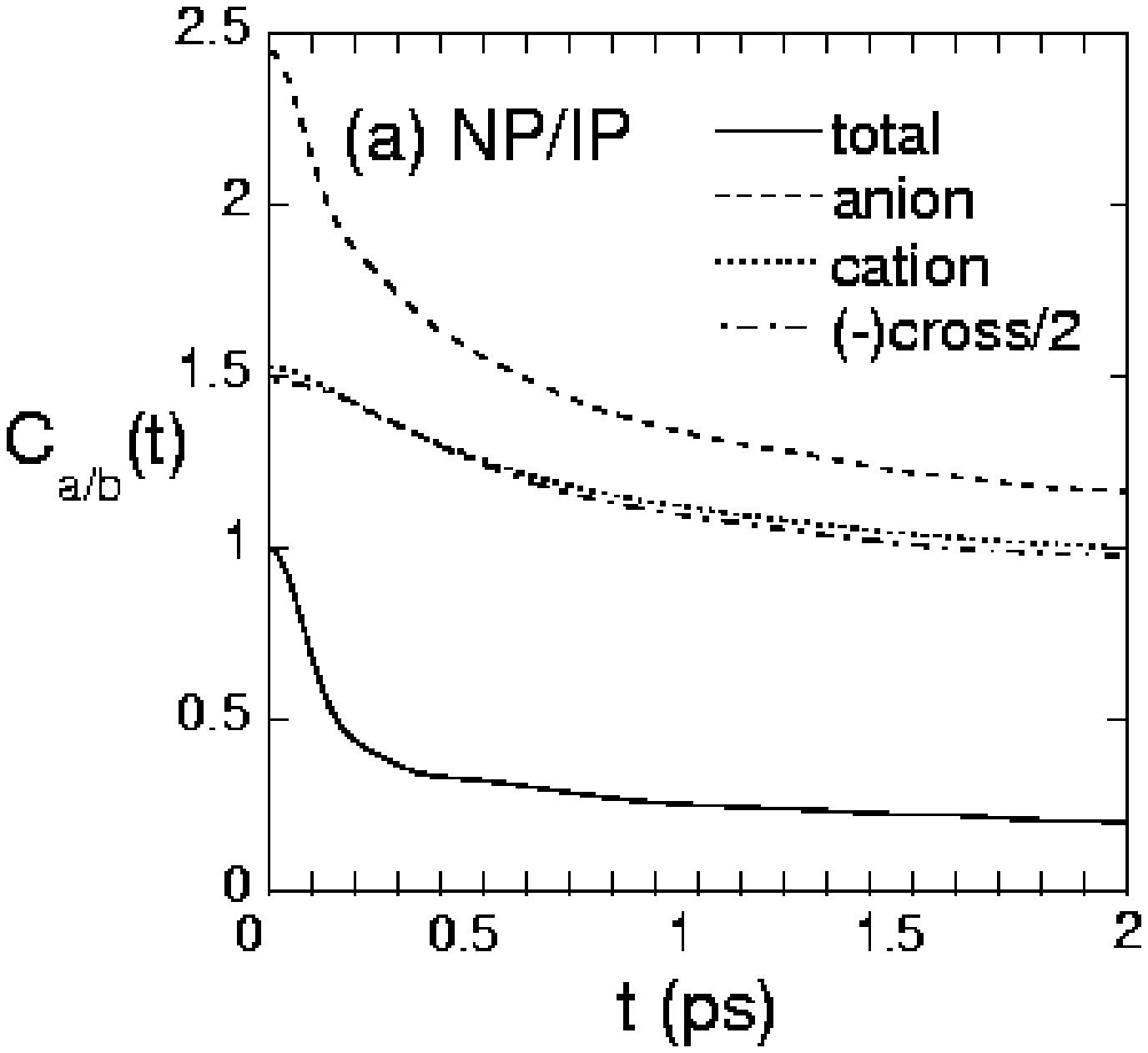} \\  
\end{minipage} 
\begin{minipage}{8.0cm} 
\centering 
\includegraphics[width=3.5in]{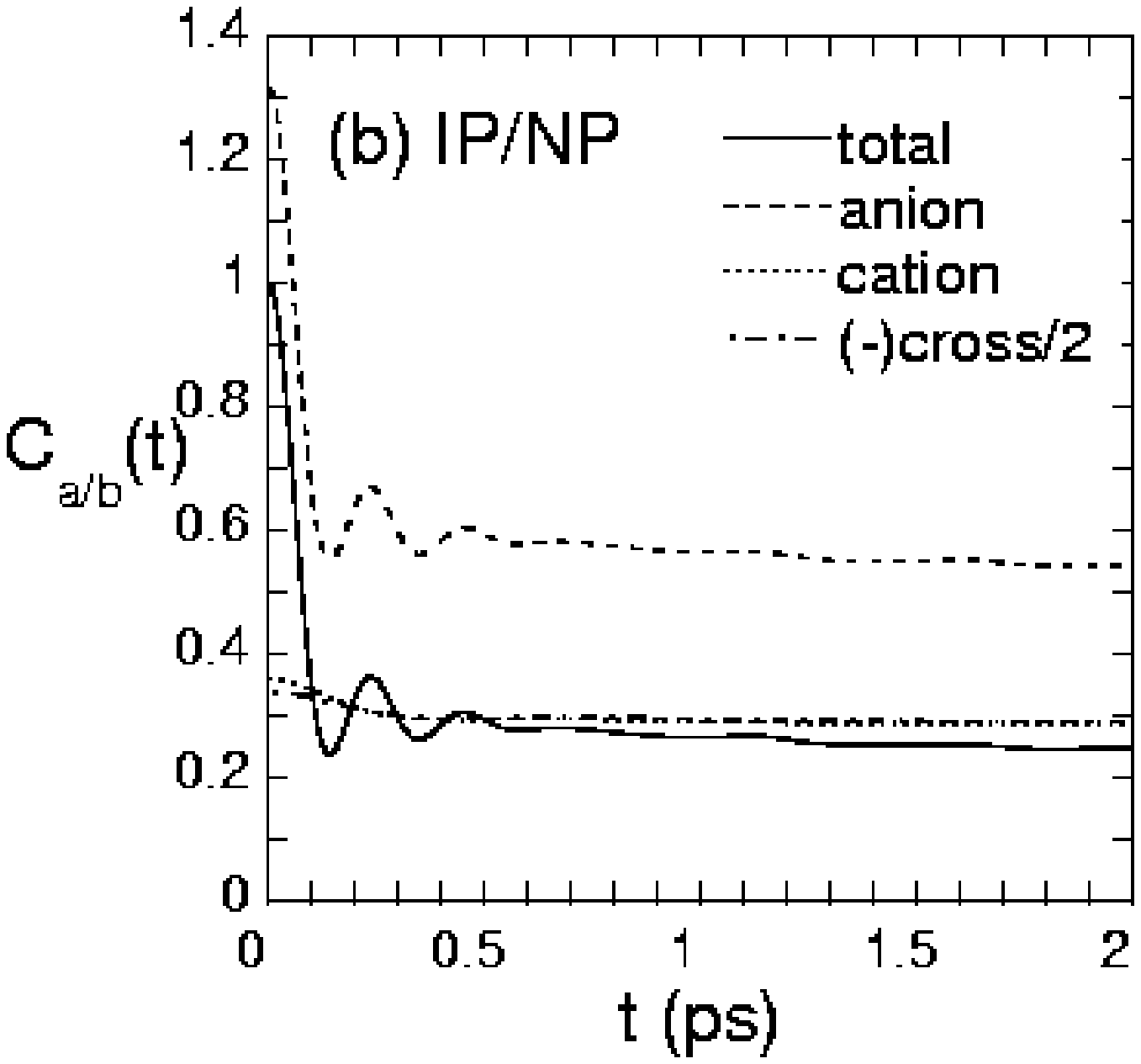} \\  
\end{minipage} \\[0.2in] 
\begin{minipage}{8.0cm} 
\centering 
\includegraphics[width=3.5in]{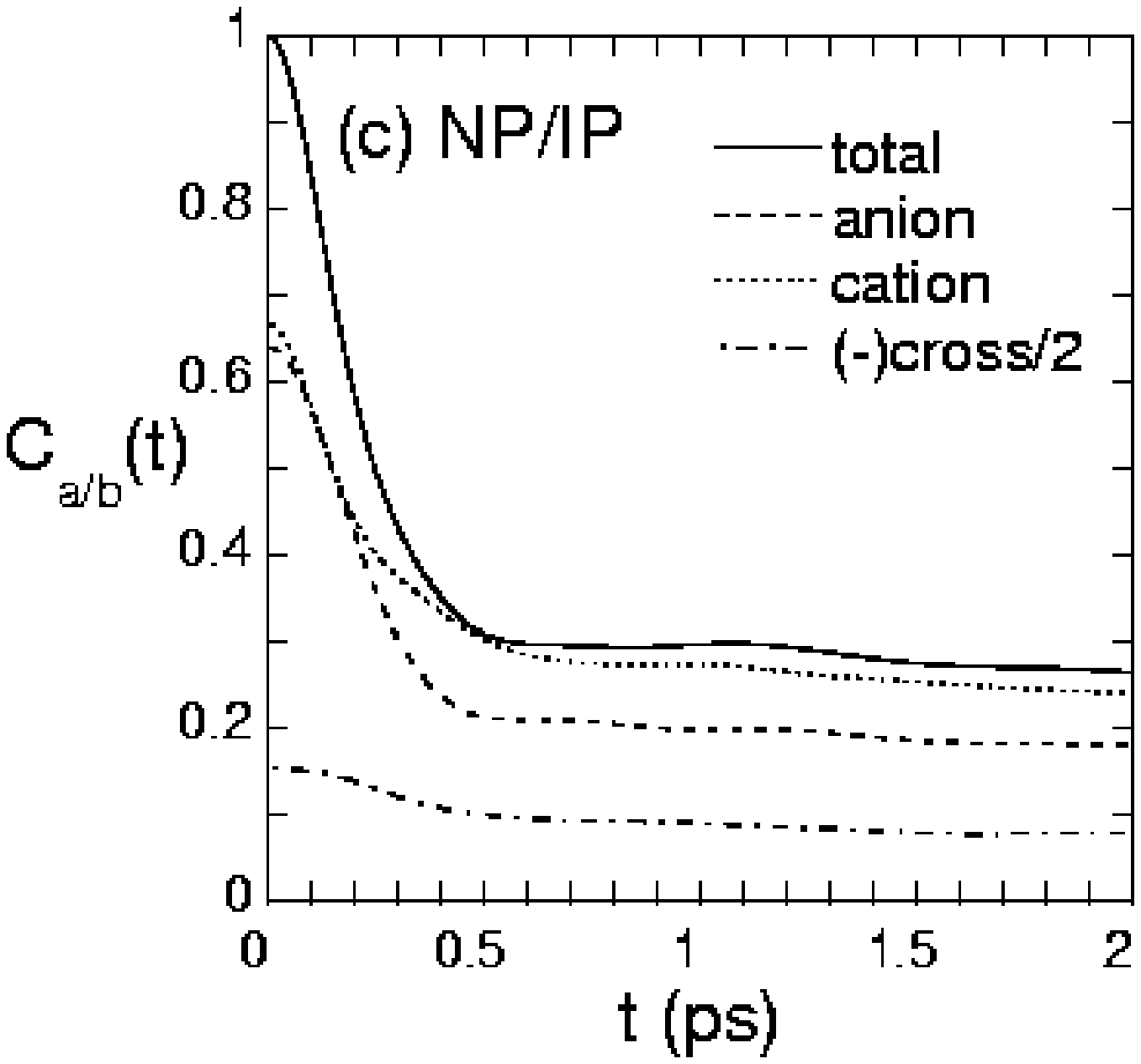} \\  
\end{minipage} 
\begin{minipage}{8.0cm} 
\centering 
\includegraphics[width=3.5in]{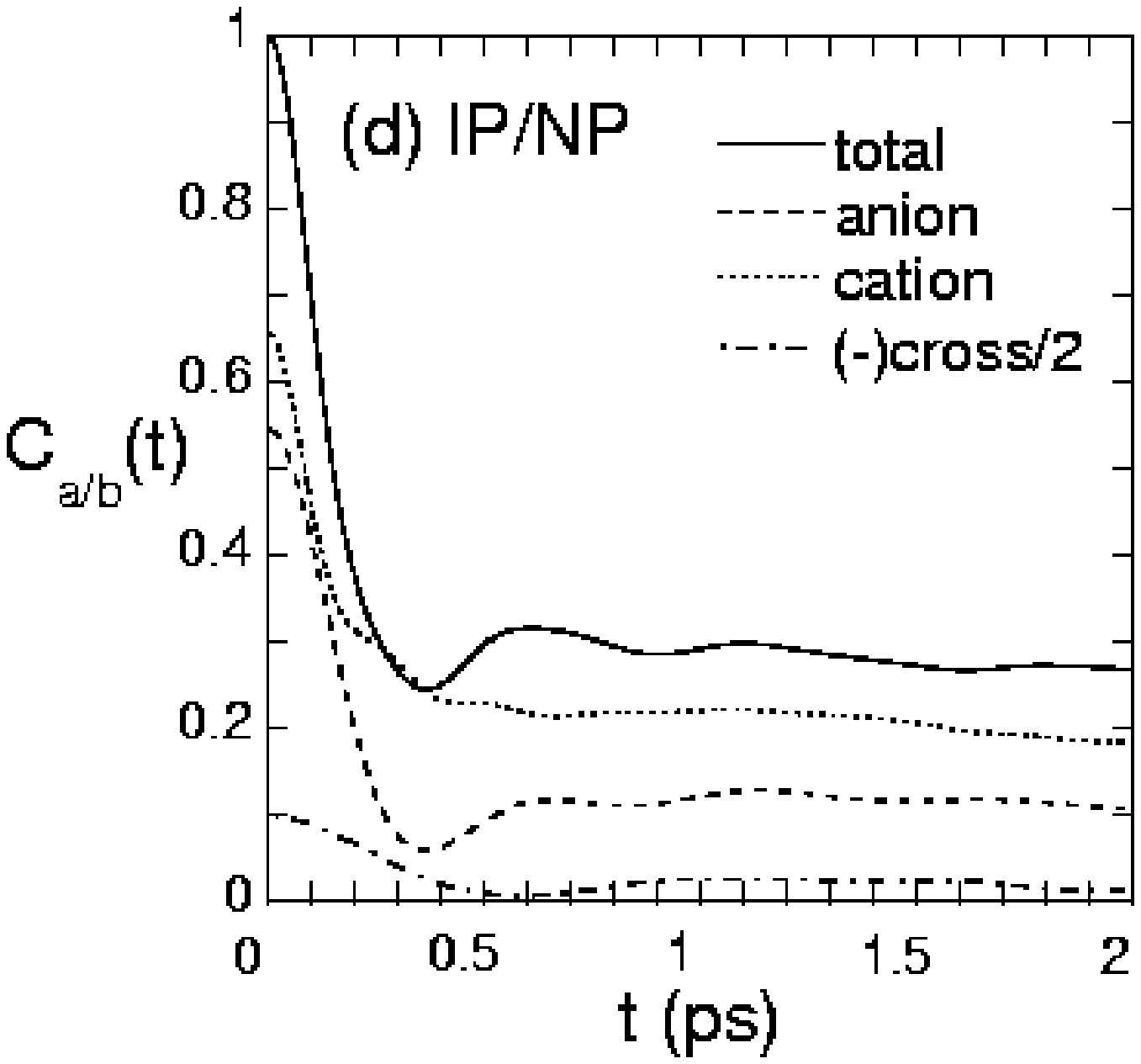}  \\  
\end{minipage} 
\vspace{10cm}
\centerline{Fig.~\ref{fig:coft:comp}}

\newpage

\centering
\includegraphics[width=4.2in]{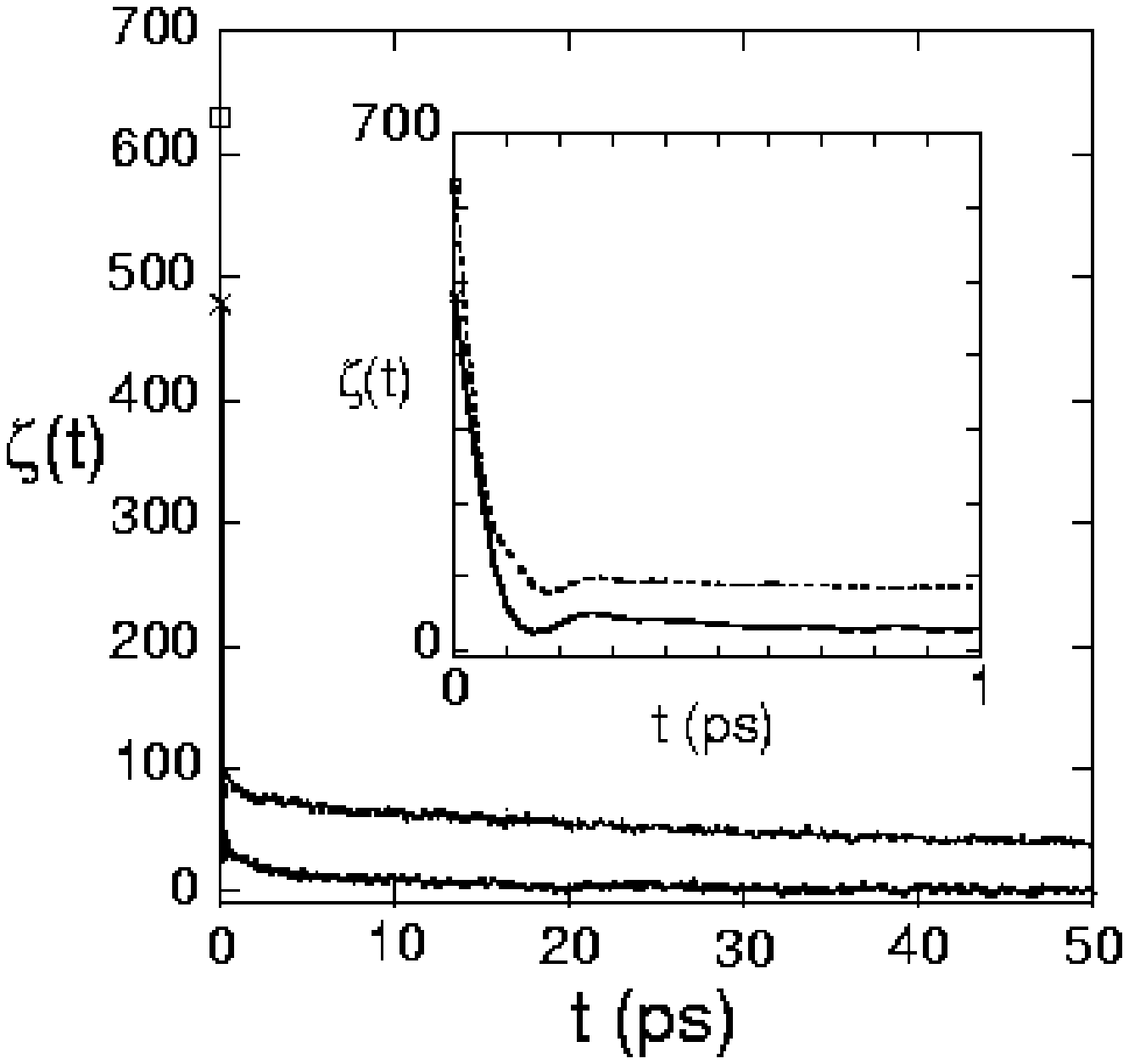}
\vspace{10cm}
\centerline{Fig.~\ref{fig:fric}}

\newpage

\centering
\includegraphics[width=4.0in]{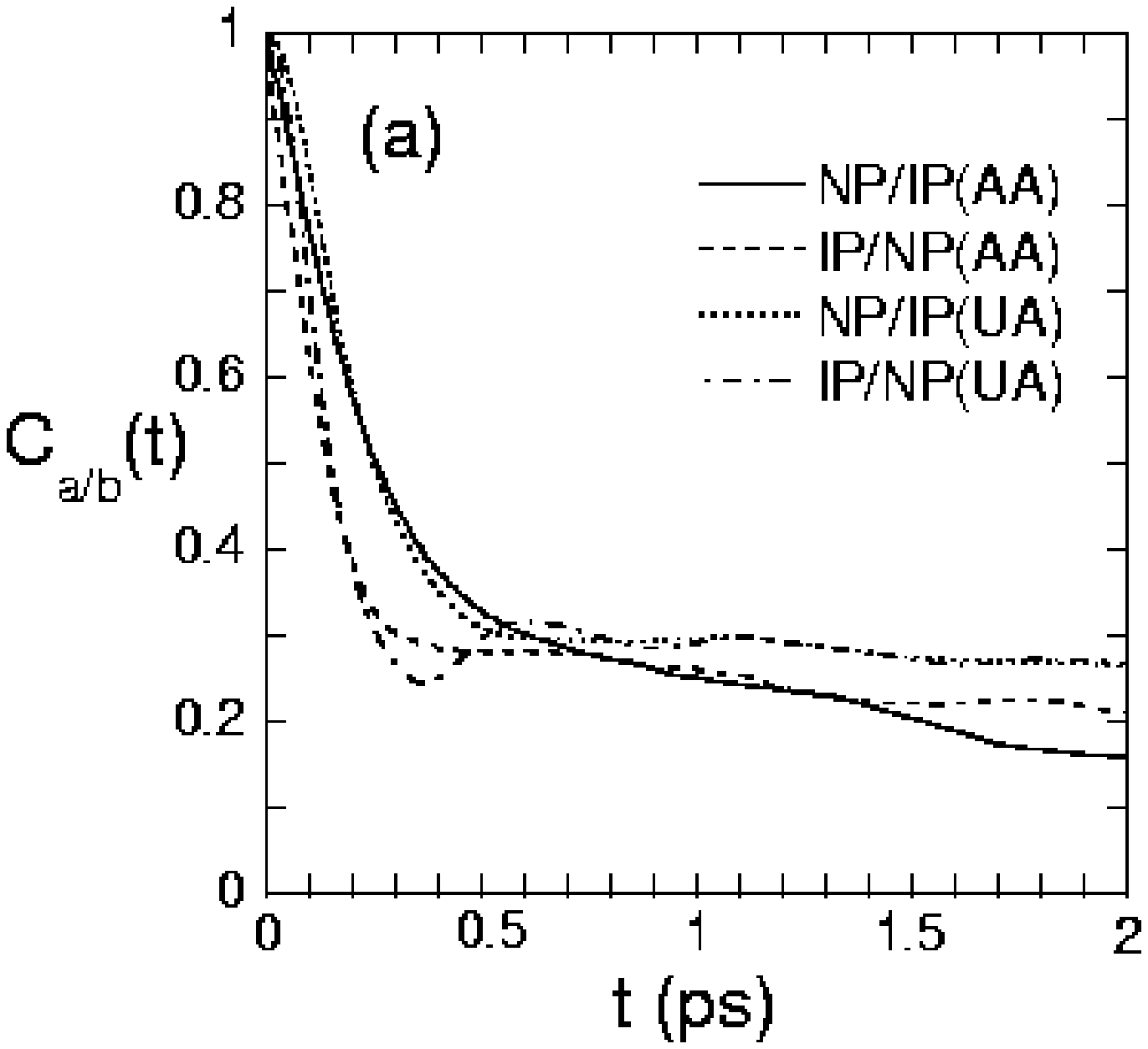}\vspace{0.3in}
\centering
\includegraphics[width=4.0in]{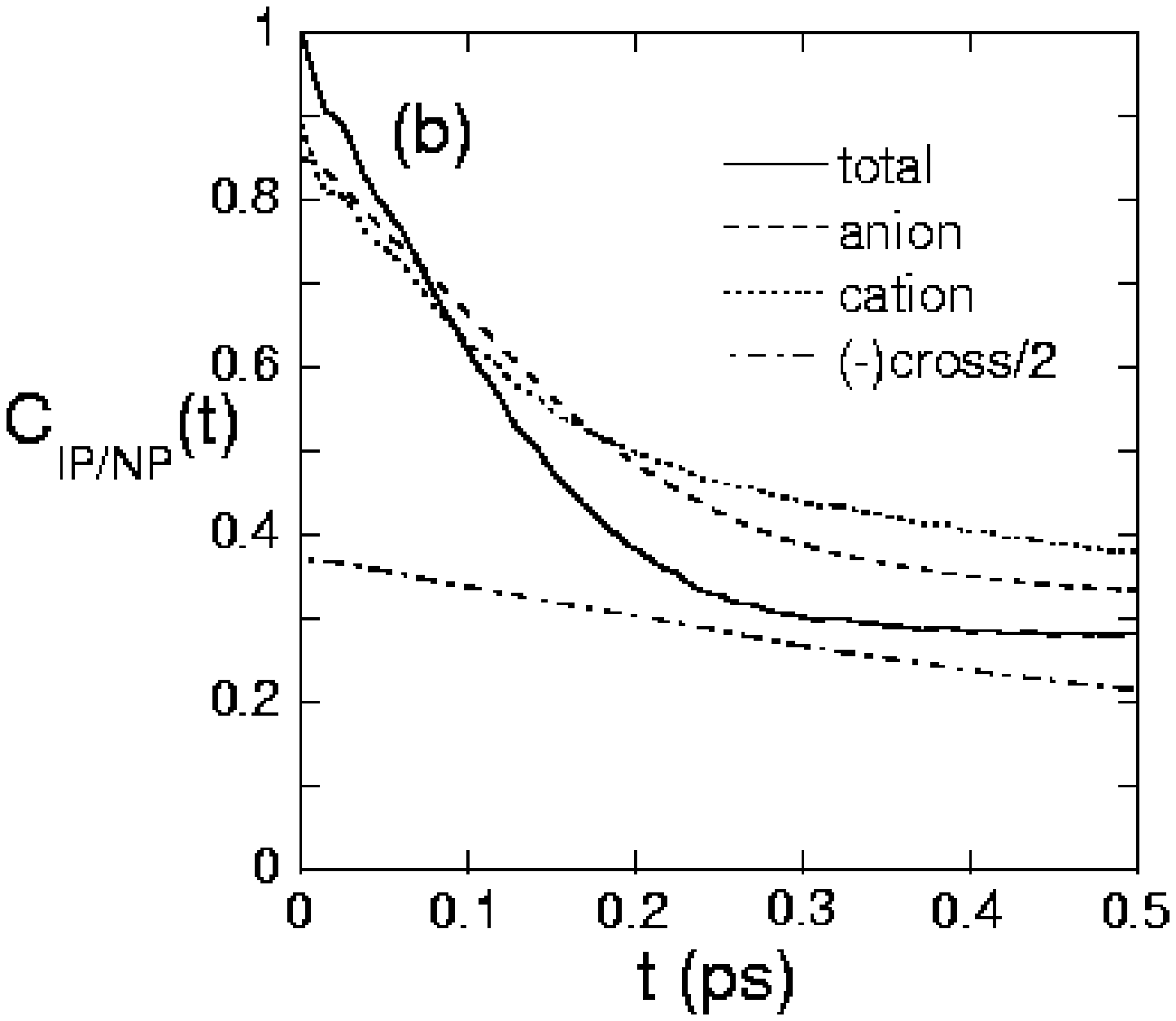}\vspace{0.3in}
\vspace{10cm}
\centerline{Fig.~\ref{fig:allatom}}

\newpage

\centering
\includegraphics[width=4.0in]{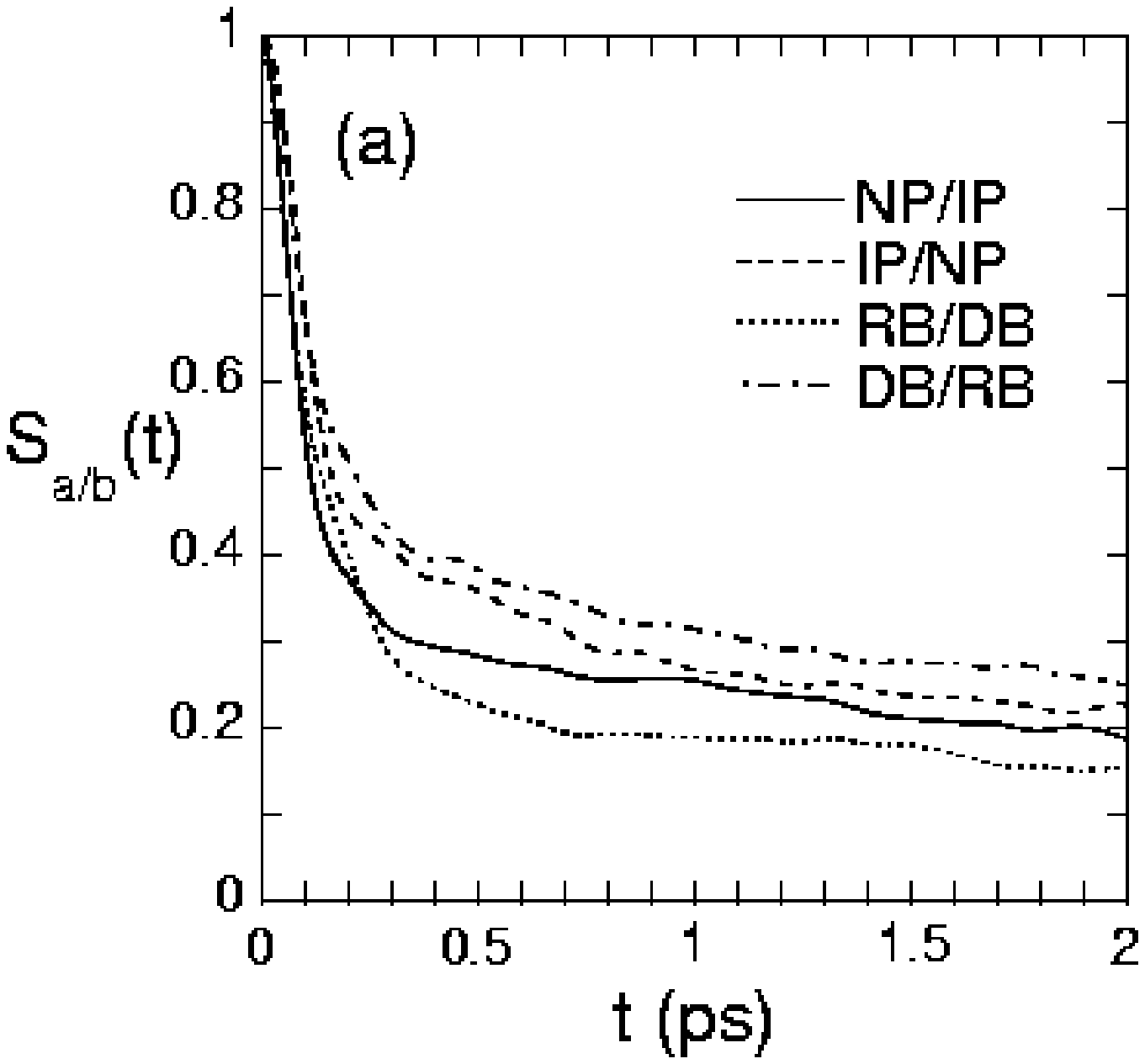}\vspace{0.3in}
\centering
\includegraphics[width=4.0in]{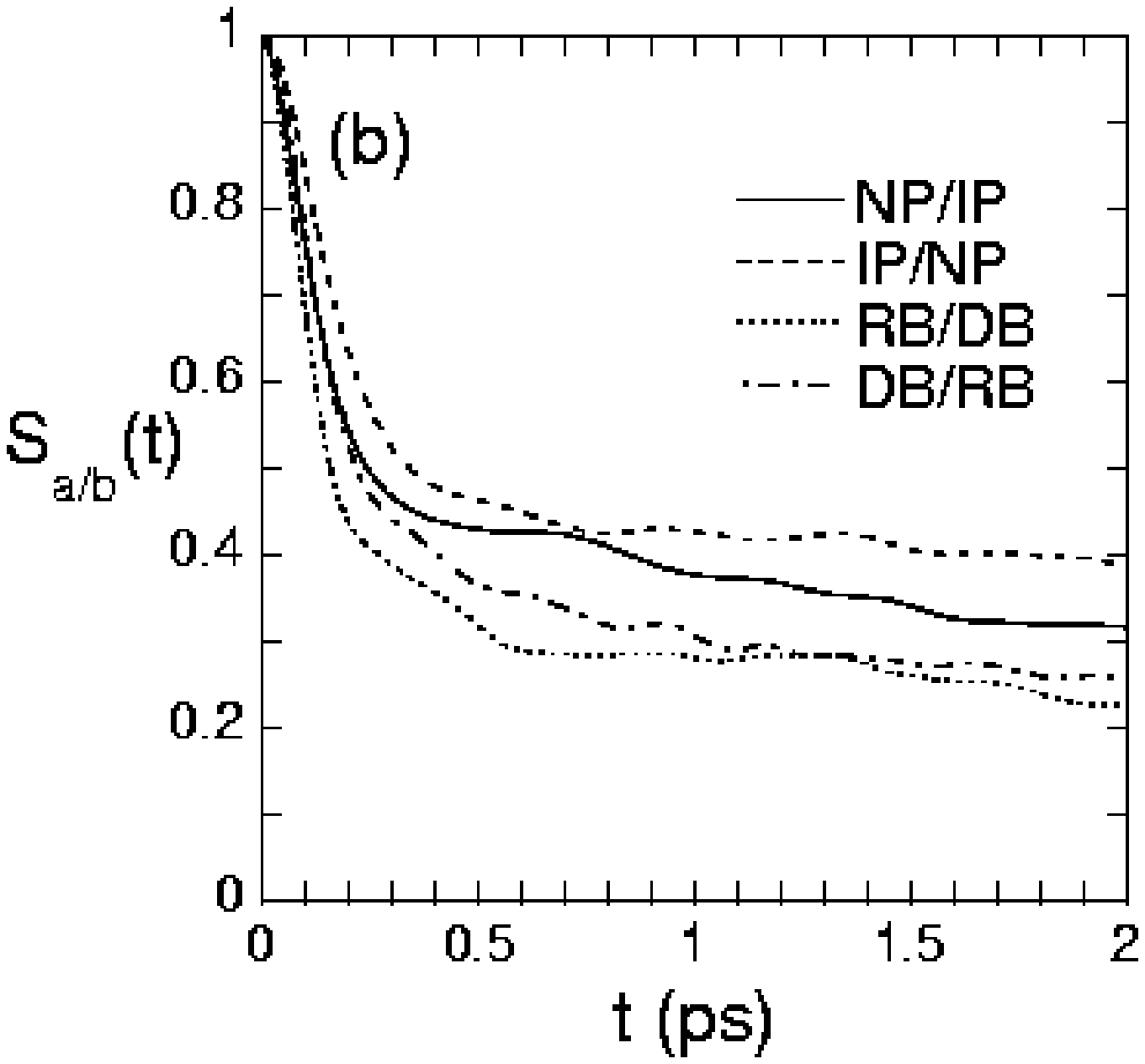}\vspace{0.3in}
\vspace{10cm}
\centerline{Fig.~\ref{fig:soft}}

\newpage

\centering 
\begin{minipage}{8.0cm} 
\centering 
\includegraphics[width=3.5in]{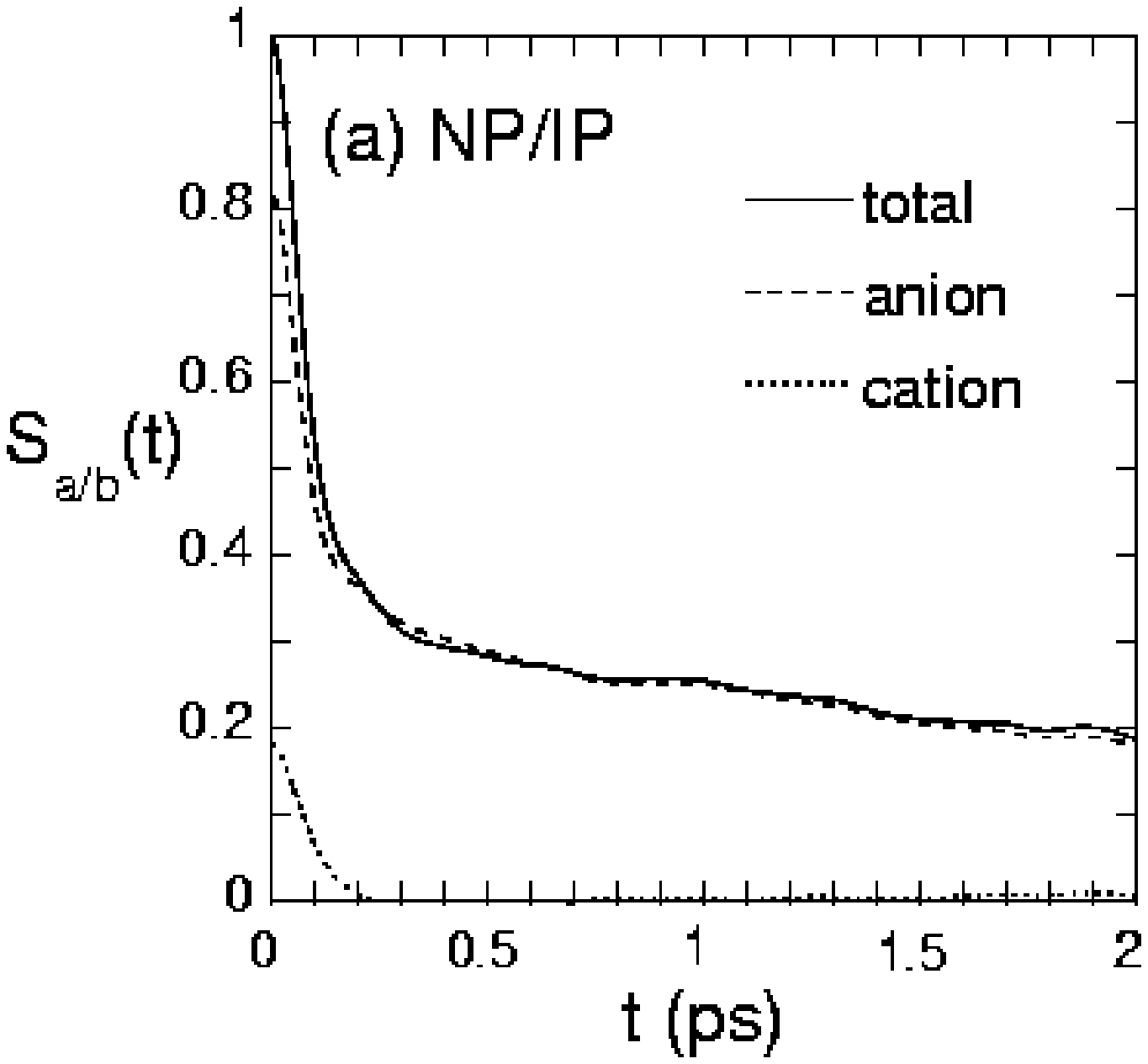} \\  
\end{minipage} 
\begin{minipage}{8.0cm} 
\centering 
\includegraphics[width=3.5in]{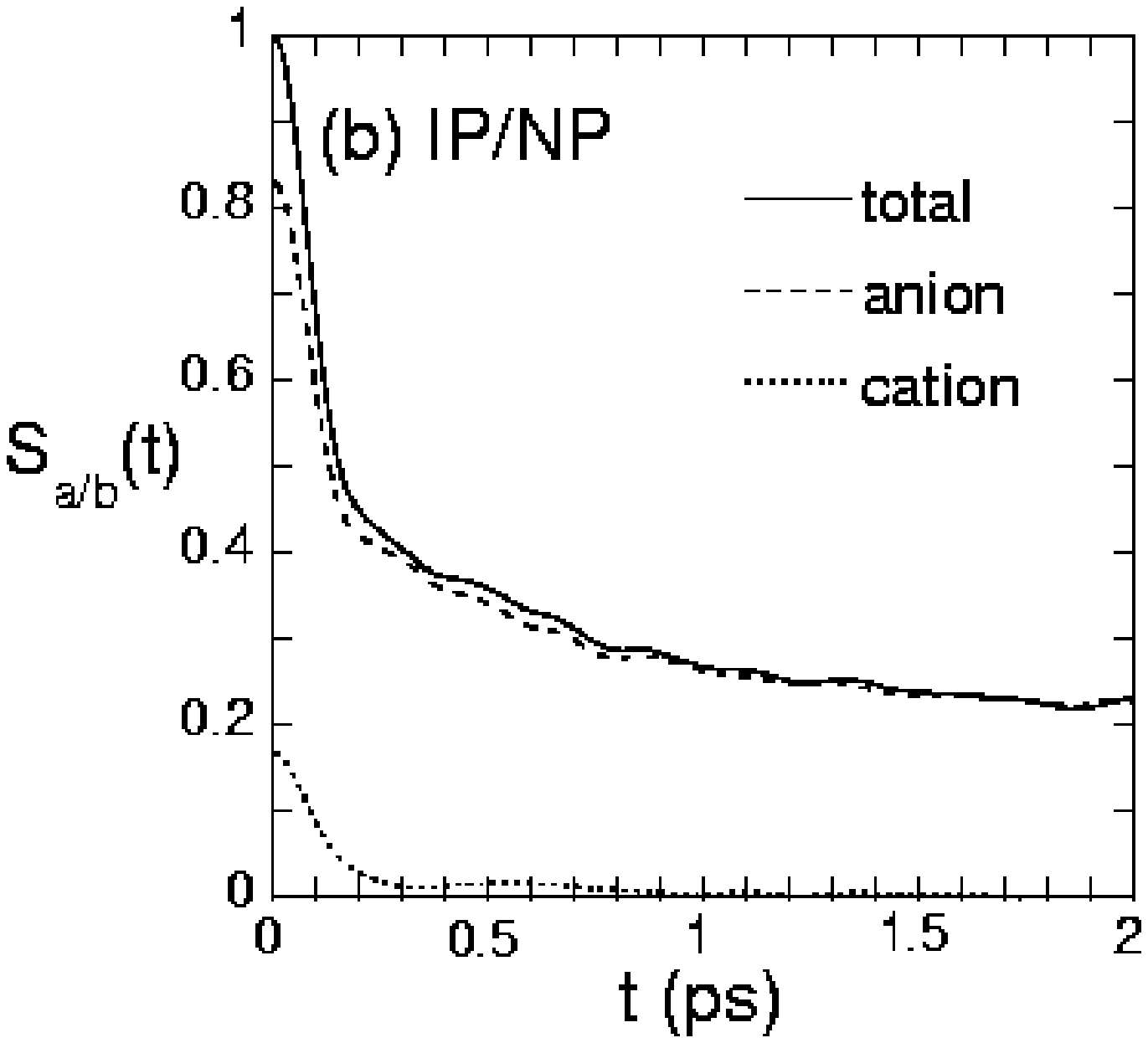} \\  
\end{minipage} \\[0.2in] 
\begin{minipage}{8.0cm} 
\centering 
\includegraphics[width=3.5in]{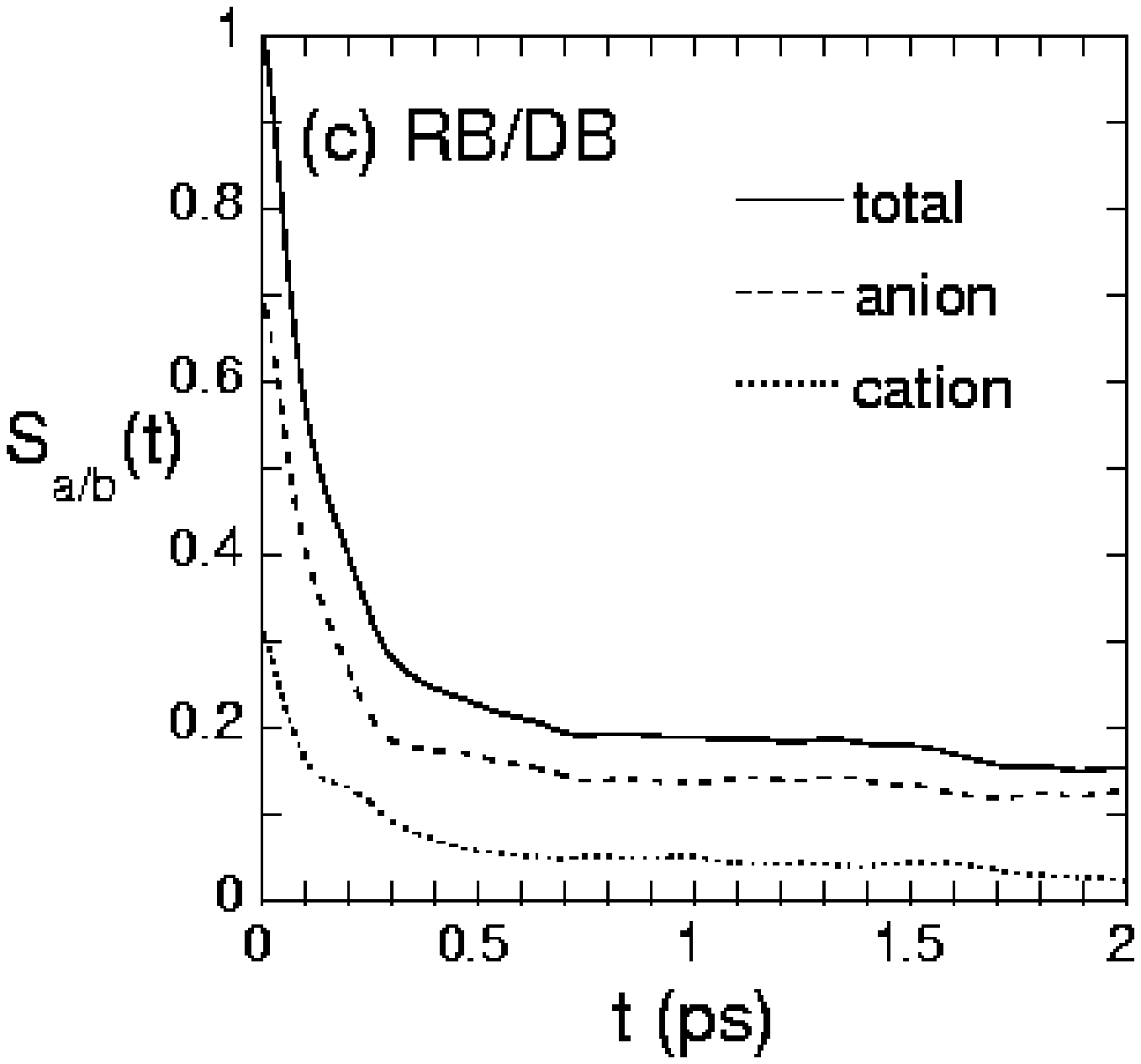} \\  
\end{minipage} 
\begin{minipage}{8.0cm} 
\centering 
\includegraphics[width=4.5in]{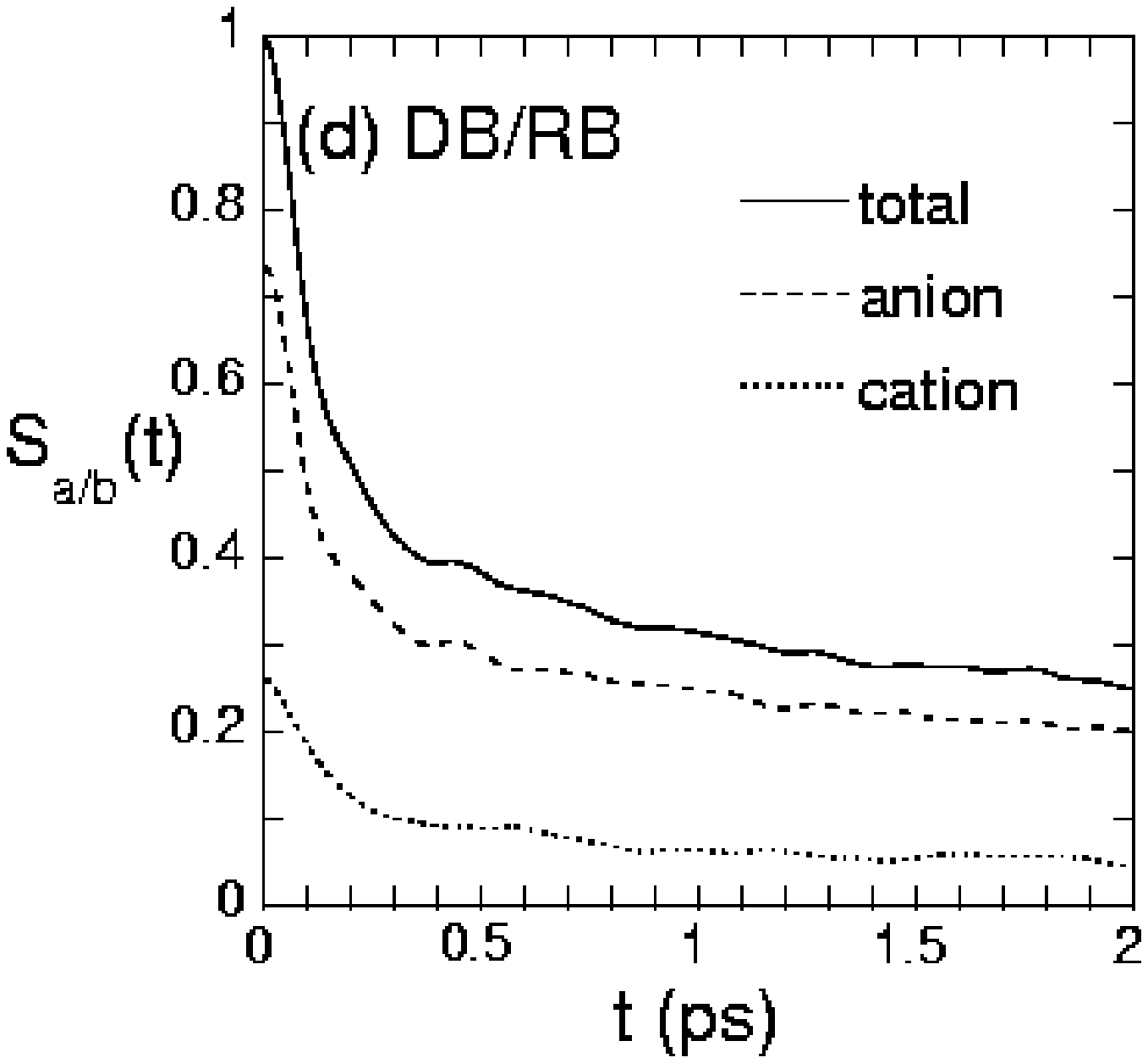}  \\  
\end{minipage} 
\vspace{10cm}
\centerline{Fig.~\ref{fig:soft:comp:emicl}}

\newpage

\centering 
\begin{minipage}{8.0cm} 
\centering 
\includegraphics[width=3.5in]{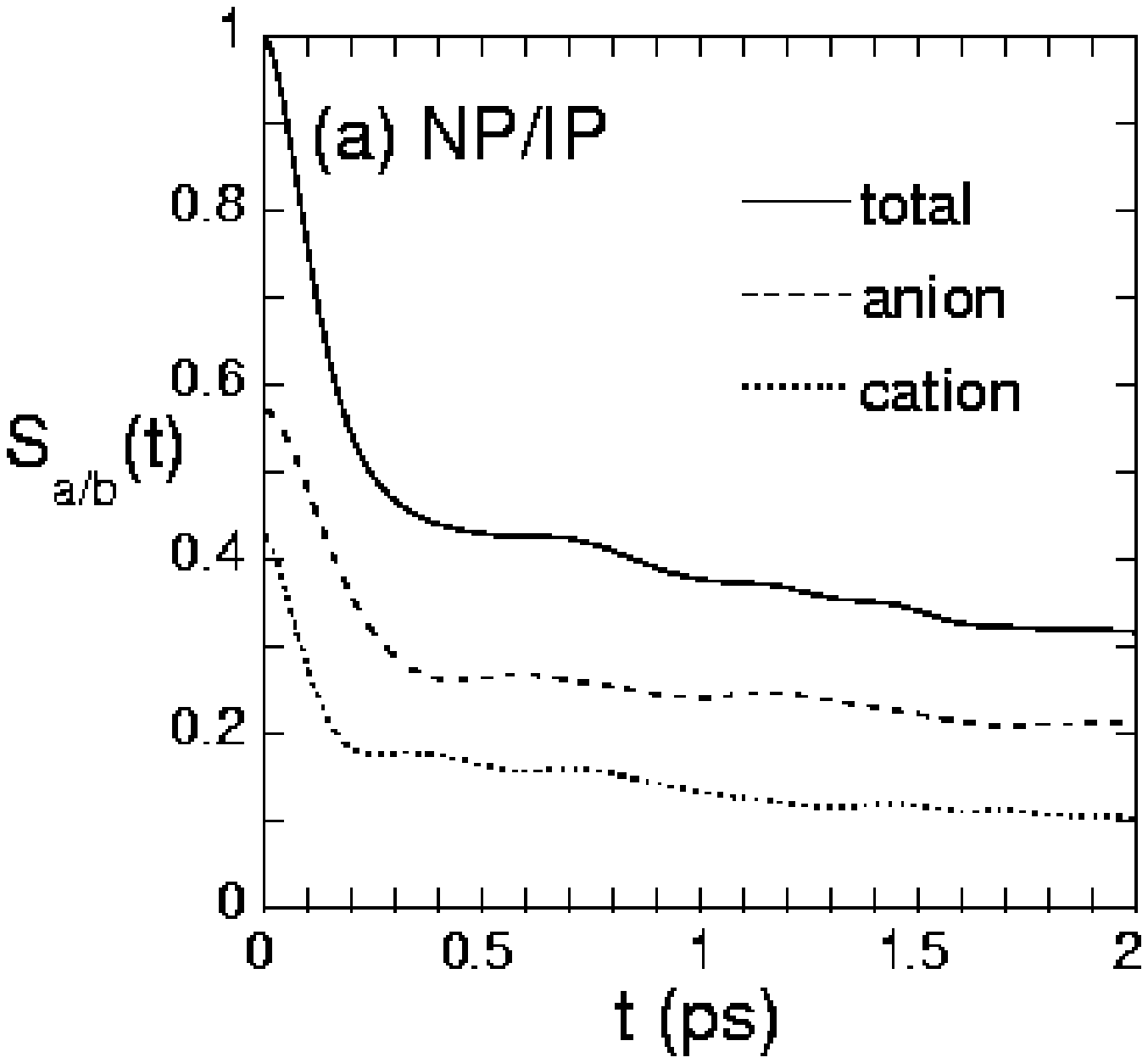} \\  
\end{minipage} 
\begin{minipage}{8.0cm} 
\centering 
\includegraphics[width=3.5in]{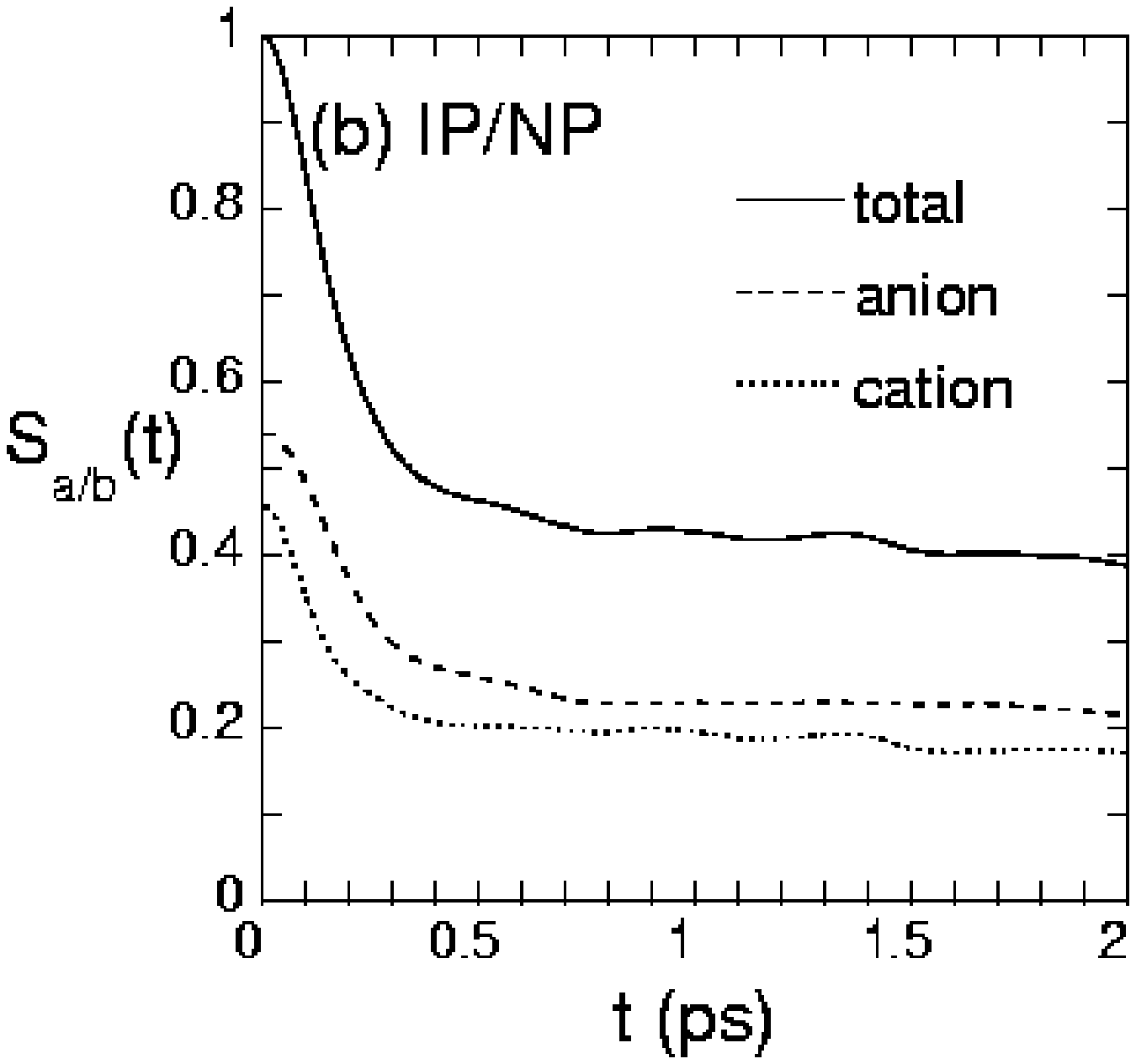} \\  
\end{minipage} \\[0.2in] 
\begin{minipage}{8.0cm} 
\centering 
\includegraphics[width=3.5in]{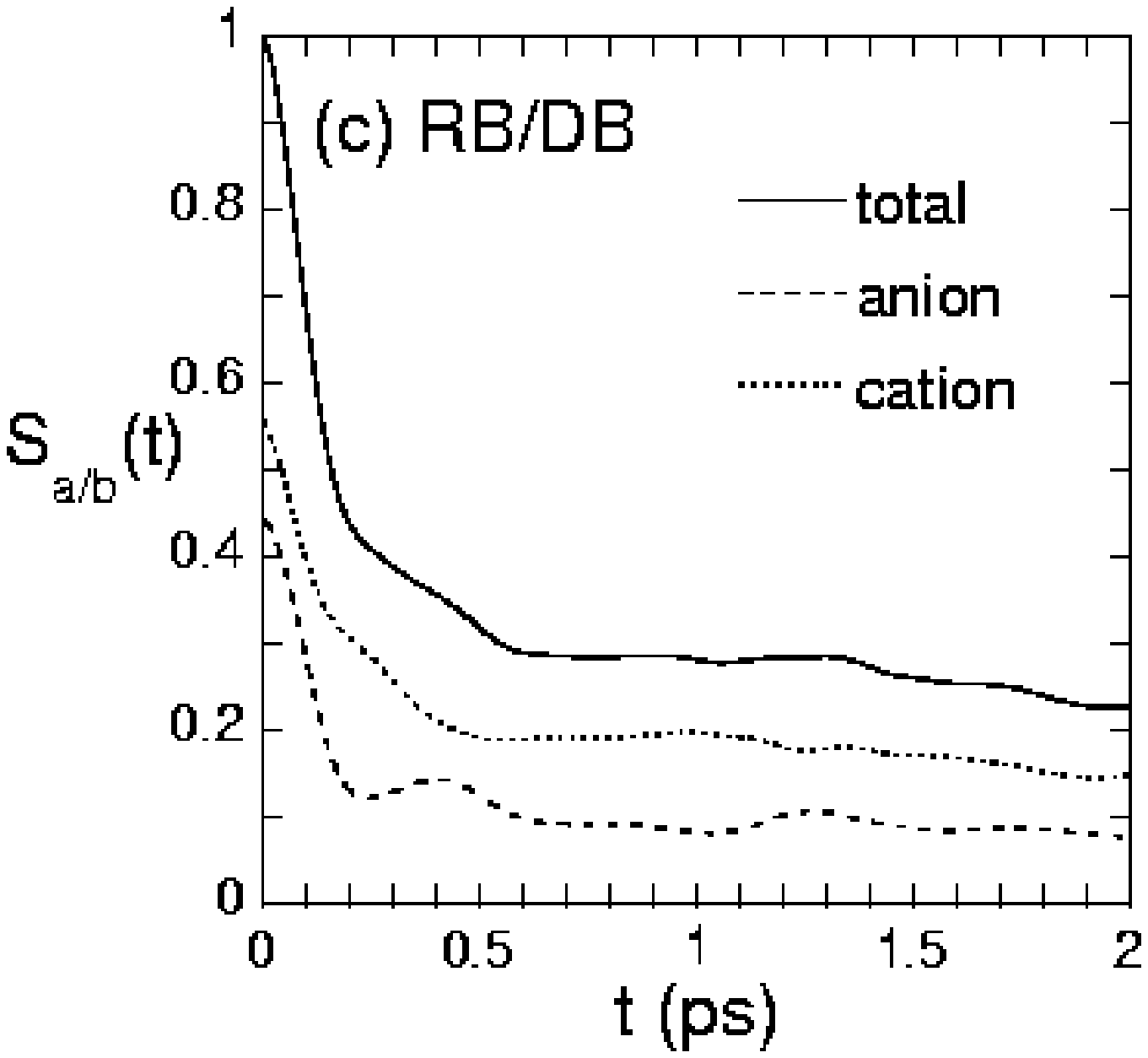} \\  
\end{minipage} 
\begin{minipage}{8.0cm} 
\centering 
\includegraphics[width=3.5in]{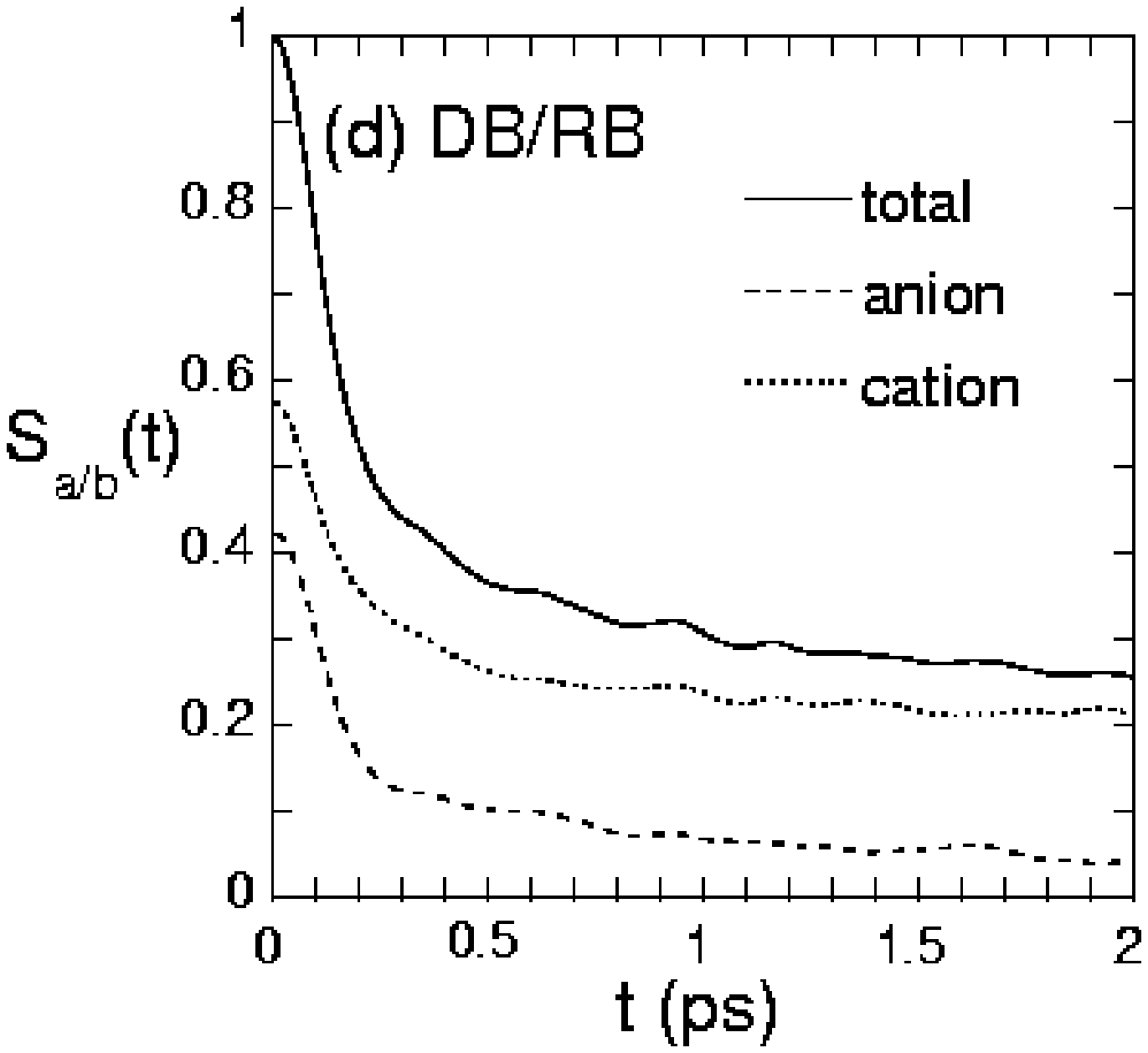}  \\  
\end{minipage} 
\vspace{10cm}
\centerline{Fig.~\ref{fig:soft:comp:emipf6}}

\newpage

\includegraphics[width=4.0in]{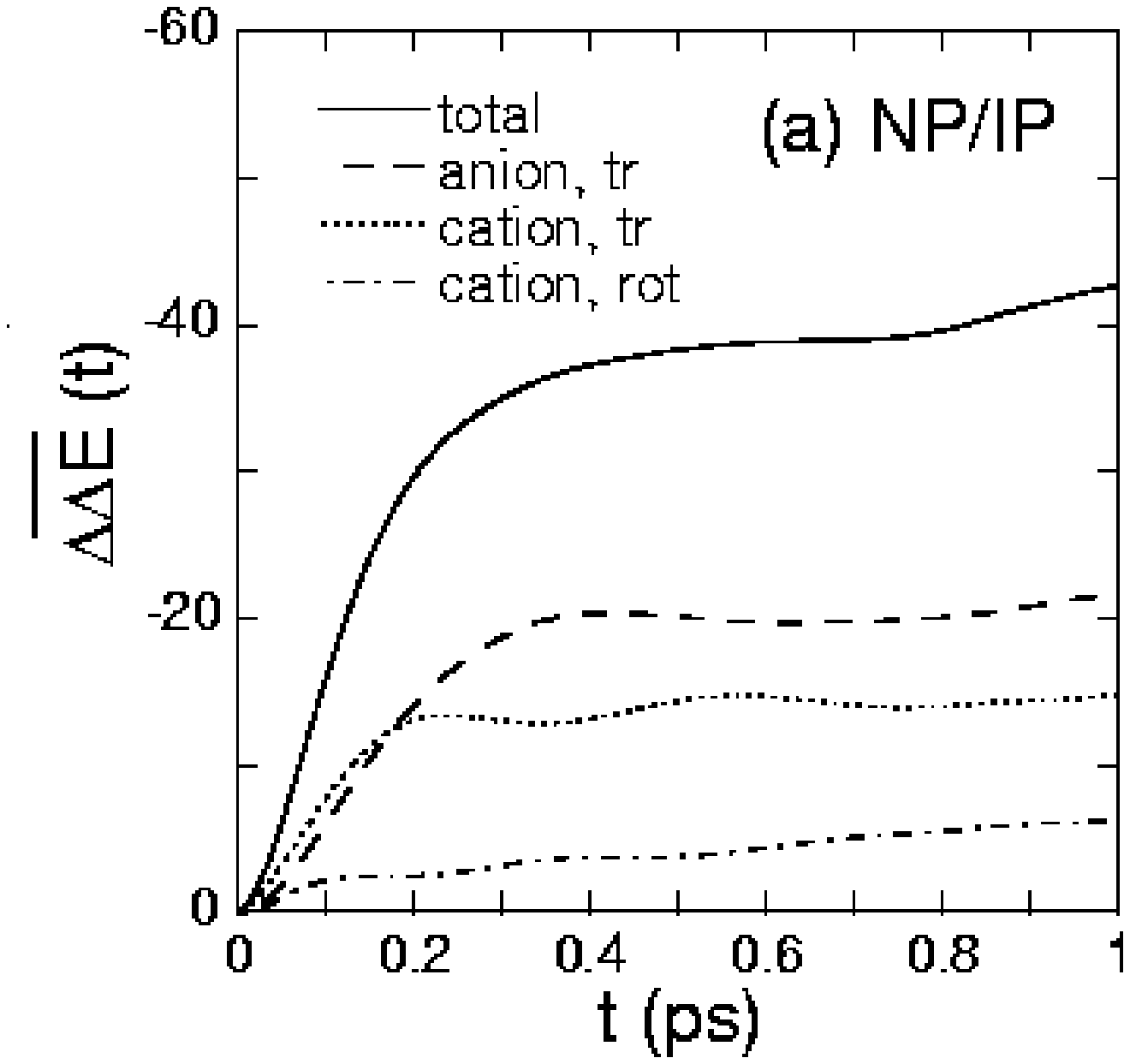}\vspace{0.3in}
\includegraphics[width=4.0in]{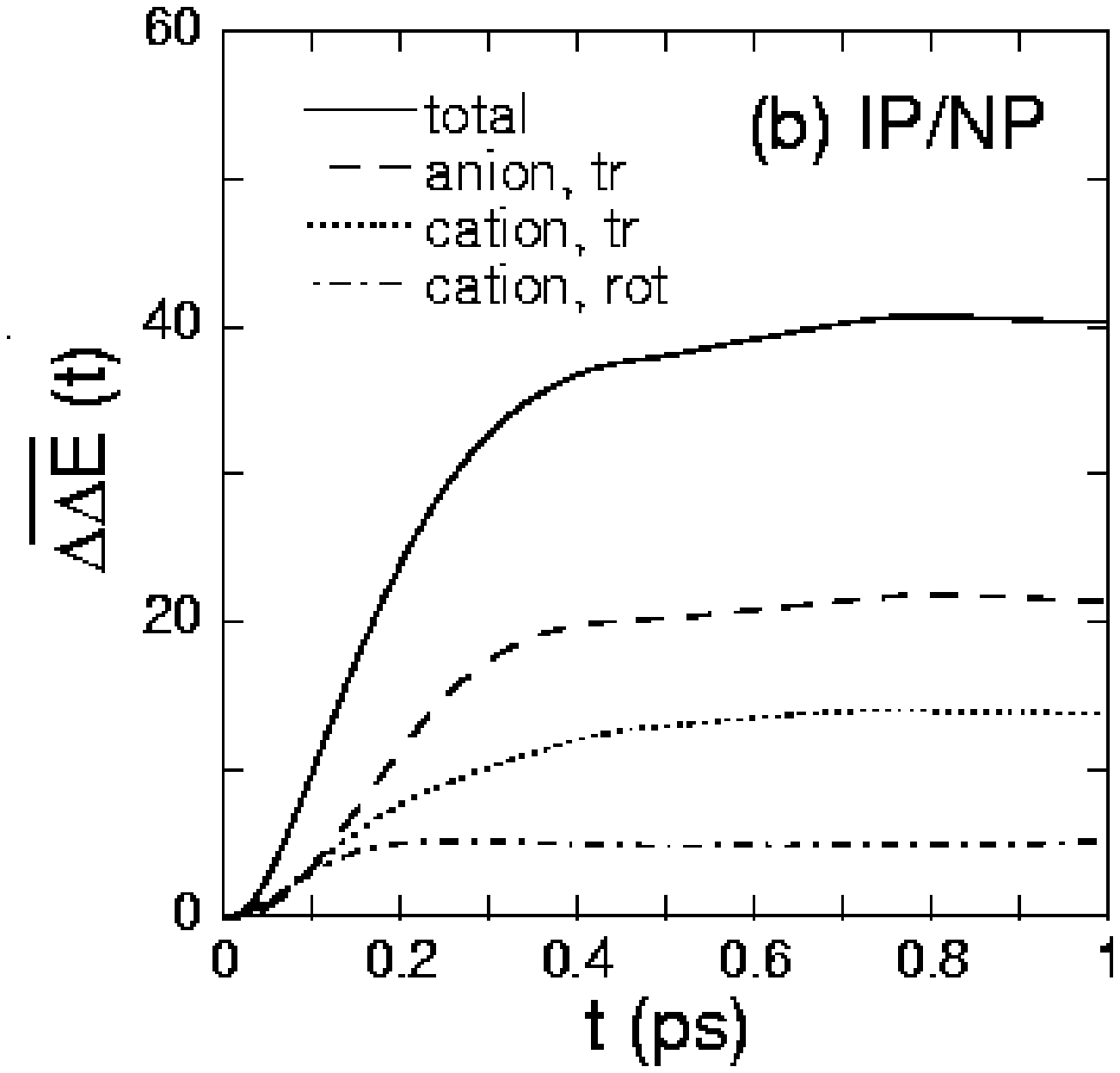}\vspace{0.3in}
\vspace{10cm}
\centerline{Fig.~\ref{fig:motion}}

\newpage

\centering 
\begin{minipage}{8.0cm} 
\centering 
\includegraphics[width=3.4in]{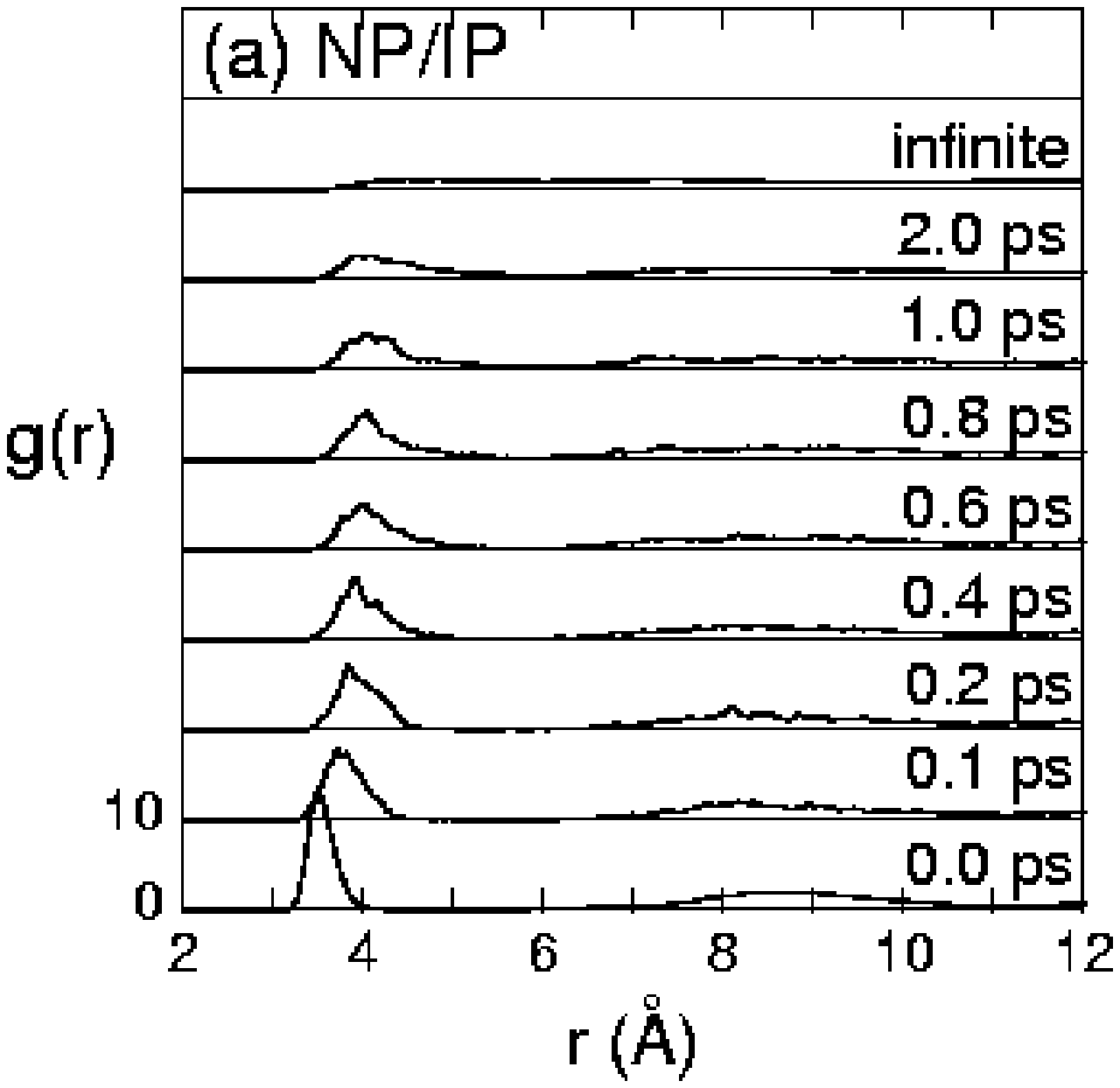} \\  
\end{minipage} 
\begin{minipage}{8.0cm} 
\centering 
\includegraphics[width=3.4in]{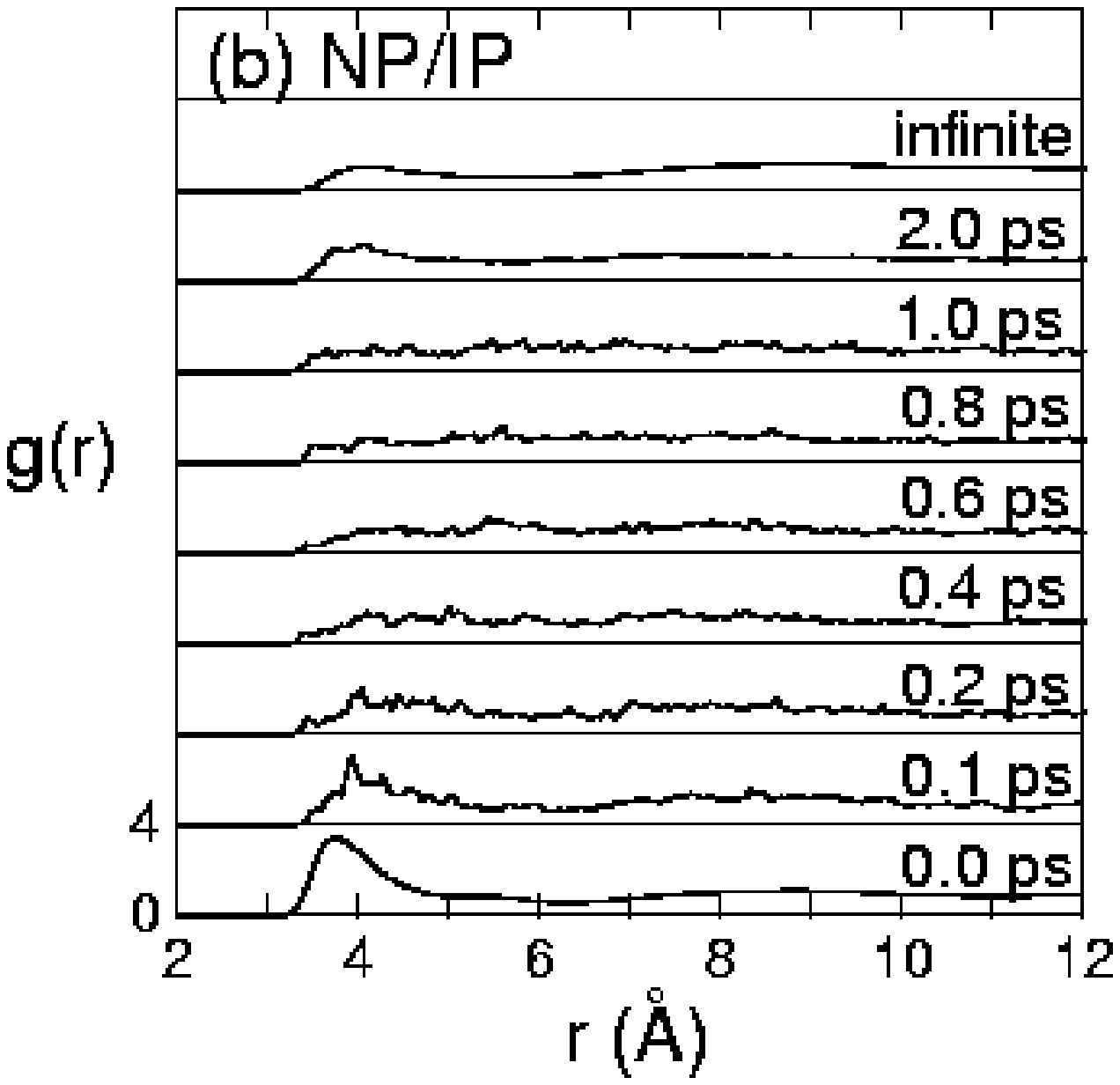} \\  
\end{minipage} \\[0.2in] 
\begin{minipage}{8.0cm} 
\centering 
\includegraphics[width=3.4in]{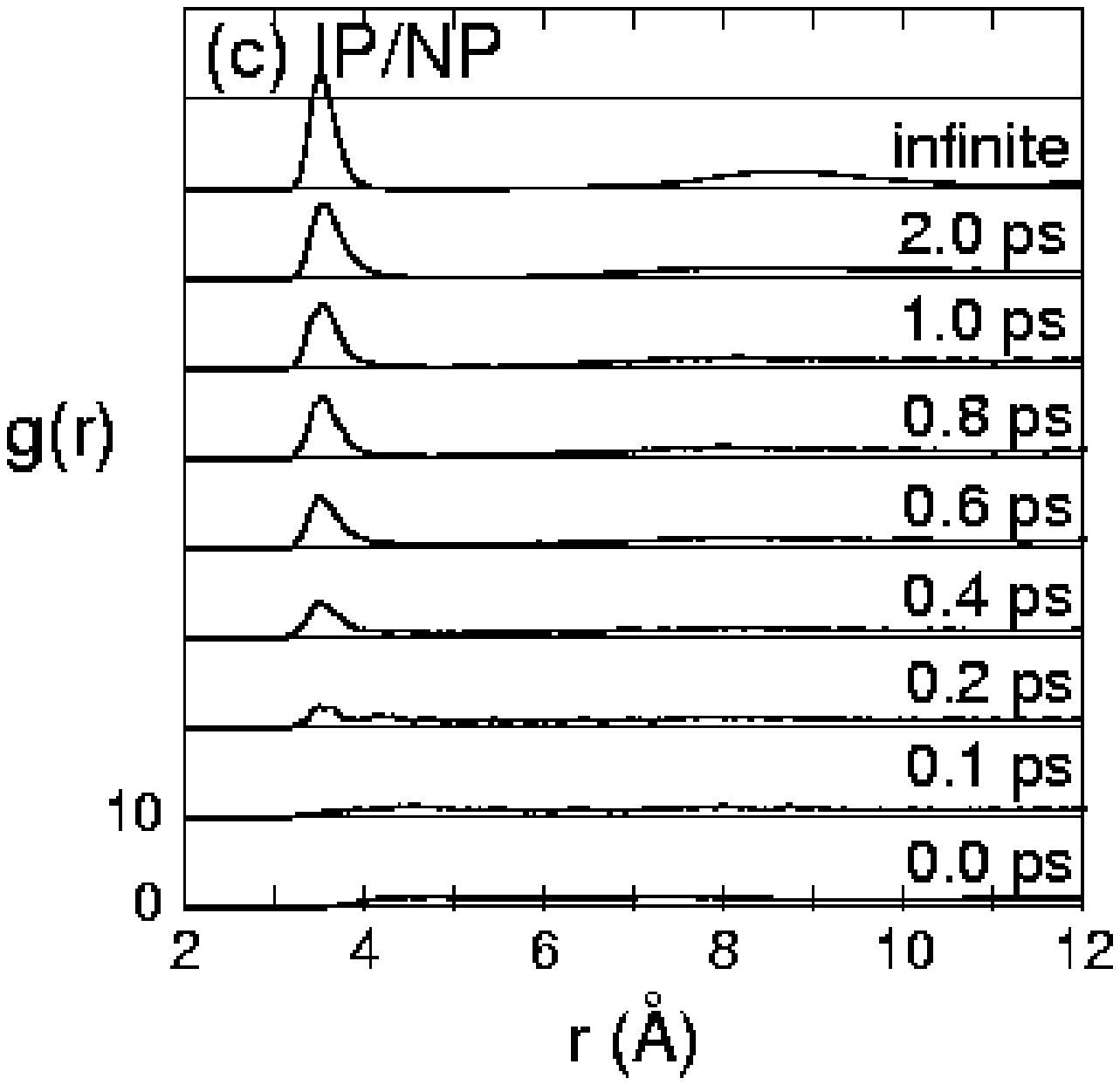} \\  
\end{minipage} 
\begin{minipage}{8.0cm} 
\centering 
\includegraphics[width=3.4in]{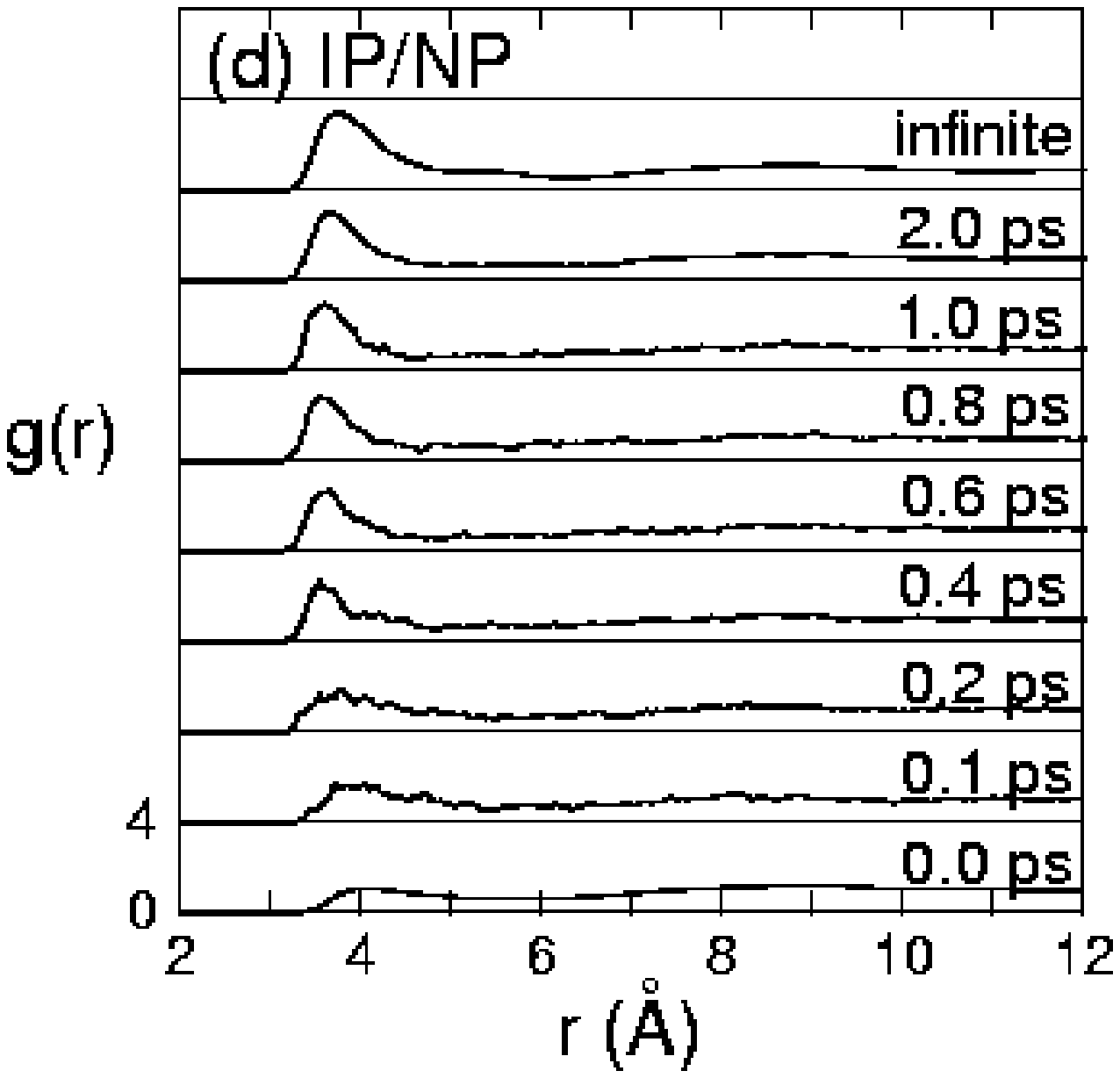}  \\  
\end{minipage} 
\vspace{10cm}
\centerline{Fig.~\ref{fig:gofrt}}

\newpage

\includegraphics[width=4.0in]{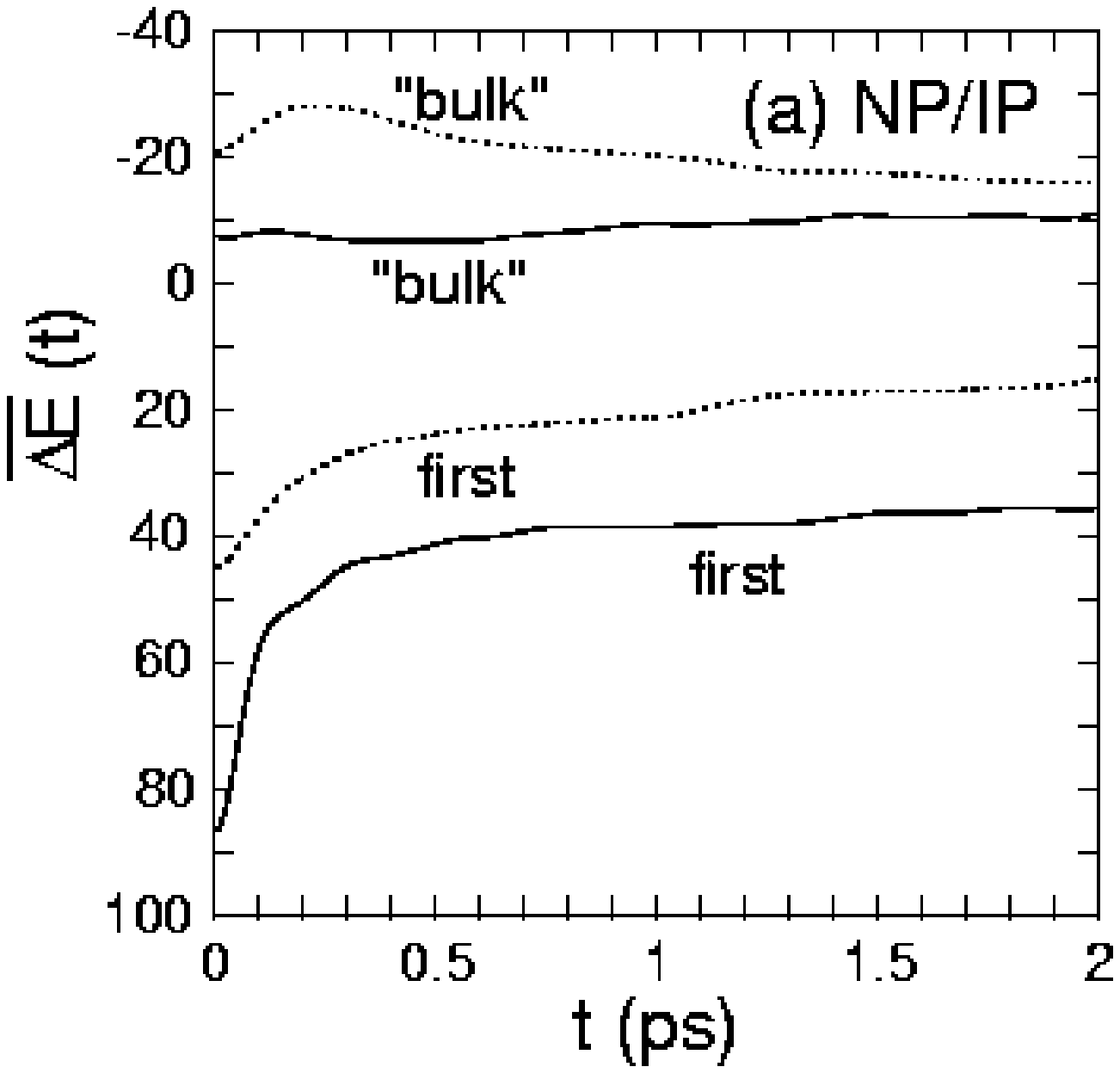}\vspace{0.3in}
\includegraphics[width=4.0in]{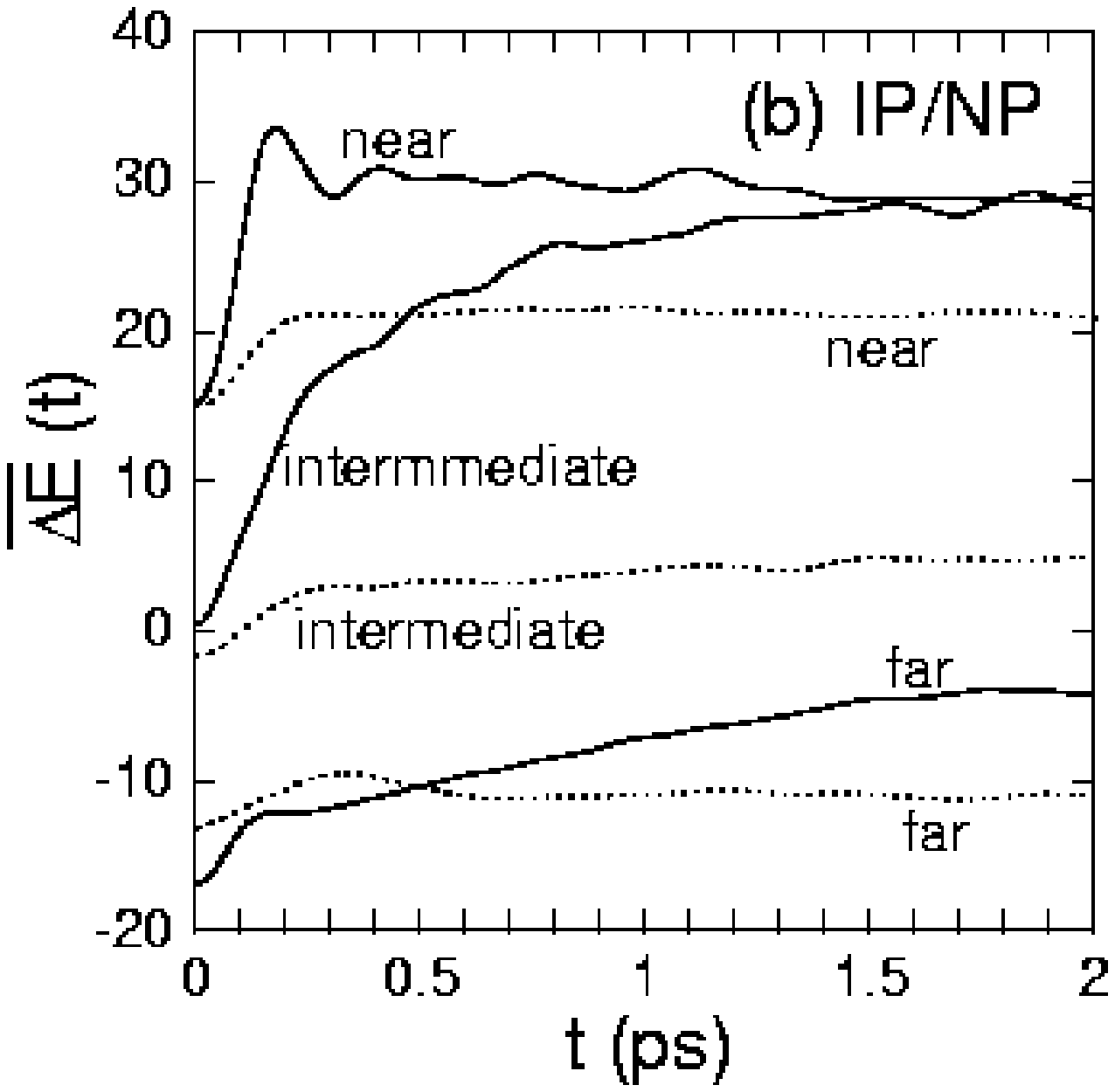}\vspace{0.3in}
\vspace{10cm}
\centerline{Fig.~\ref{fig:deltae:ion}}

\newpage

\includegraphics[width=4.0in]{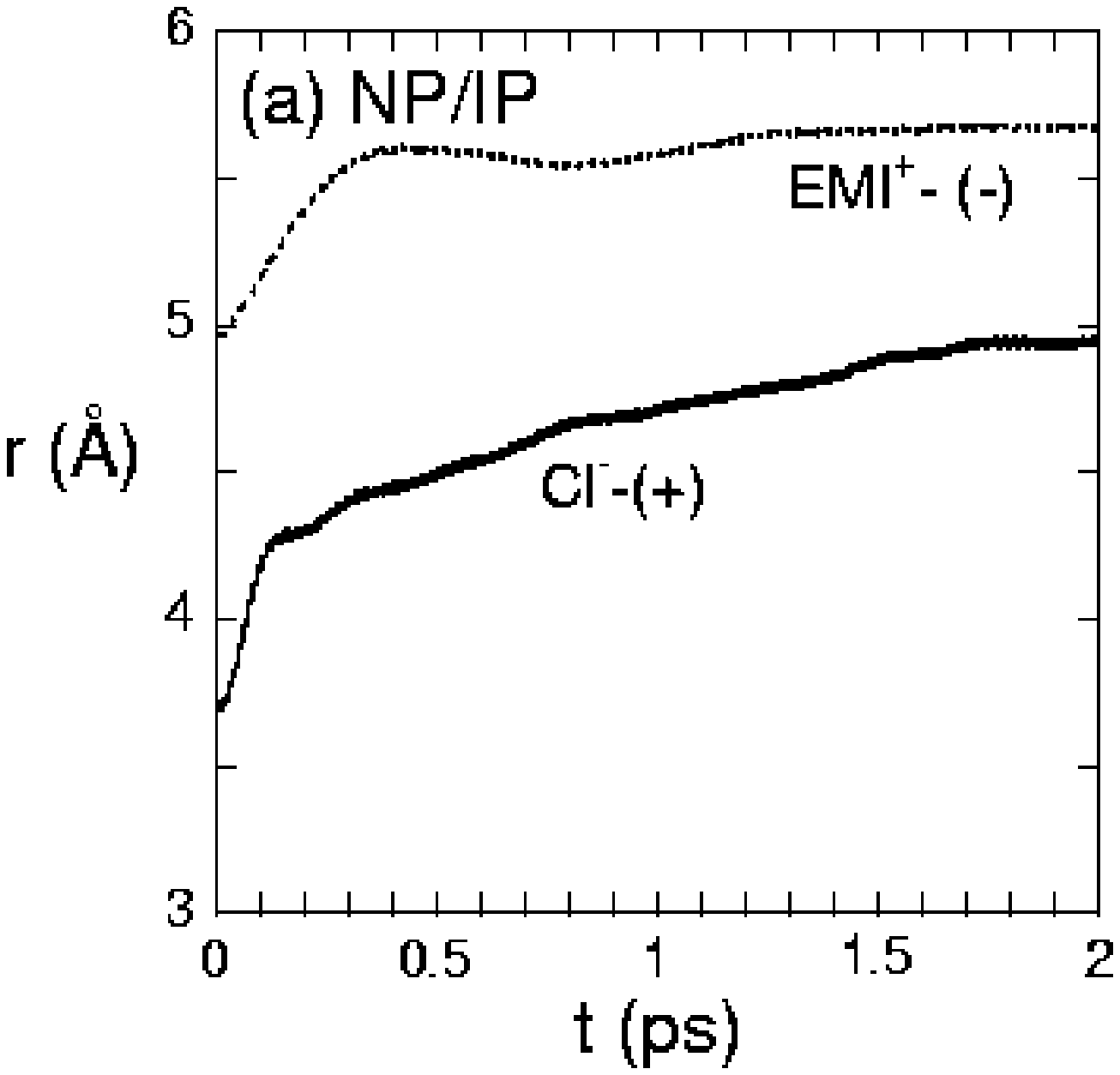}\vspace{0.3in}
\includegraphics[width=4.0in]{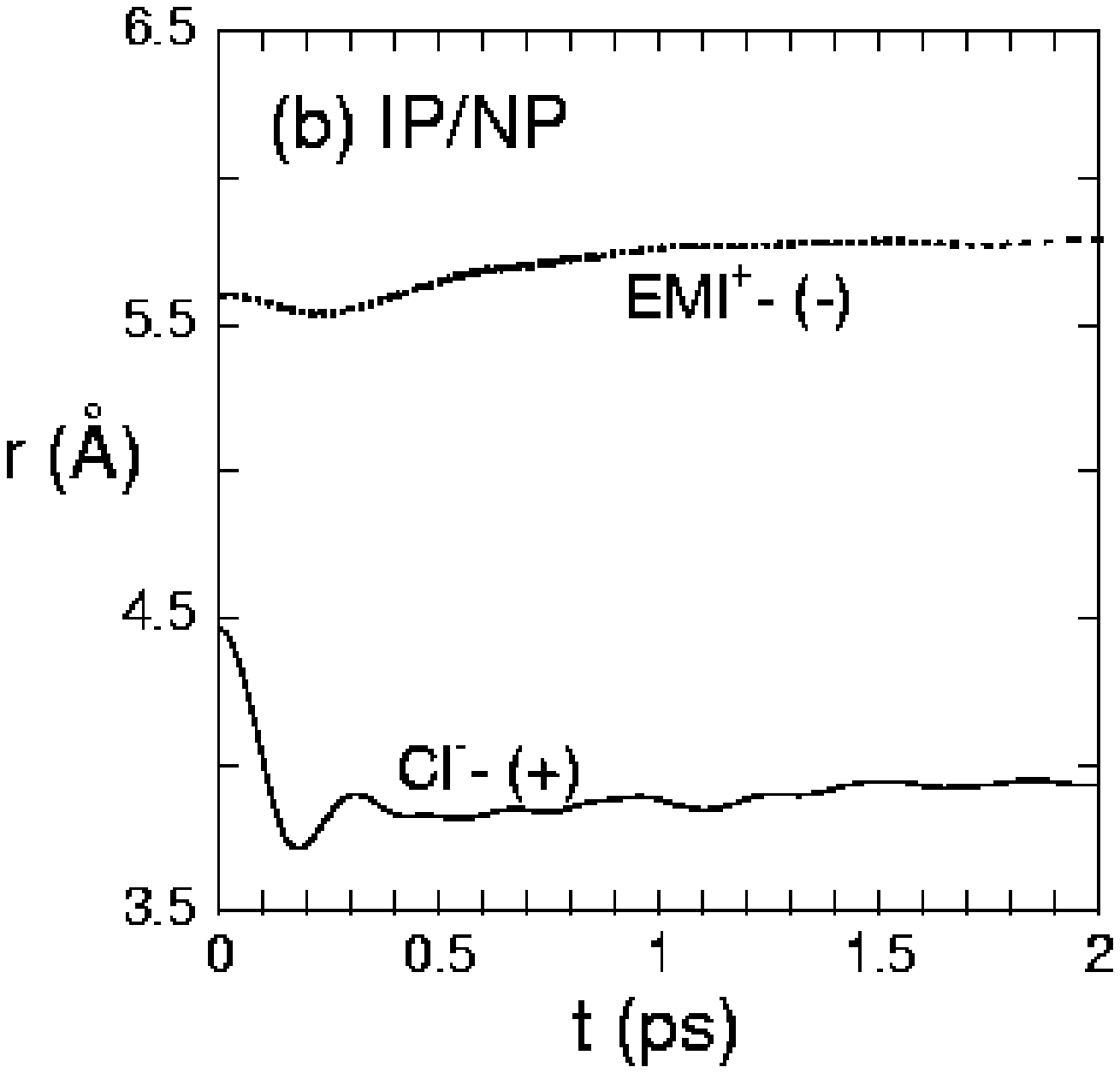}\vspace{0.3in}
\vspace{10cm}
\centerline{Fig.~\ref{fig:distance}}

\newpage

\includegraphics[width=4.0in]{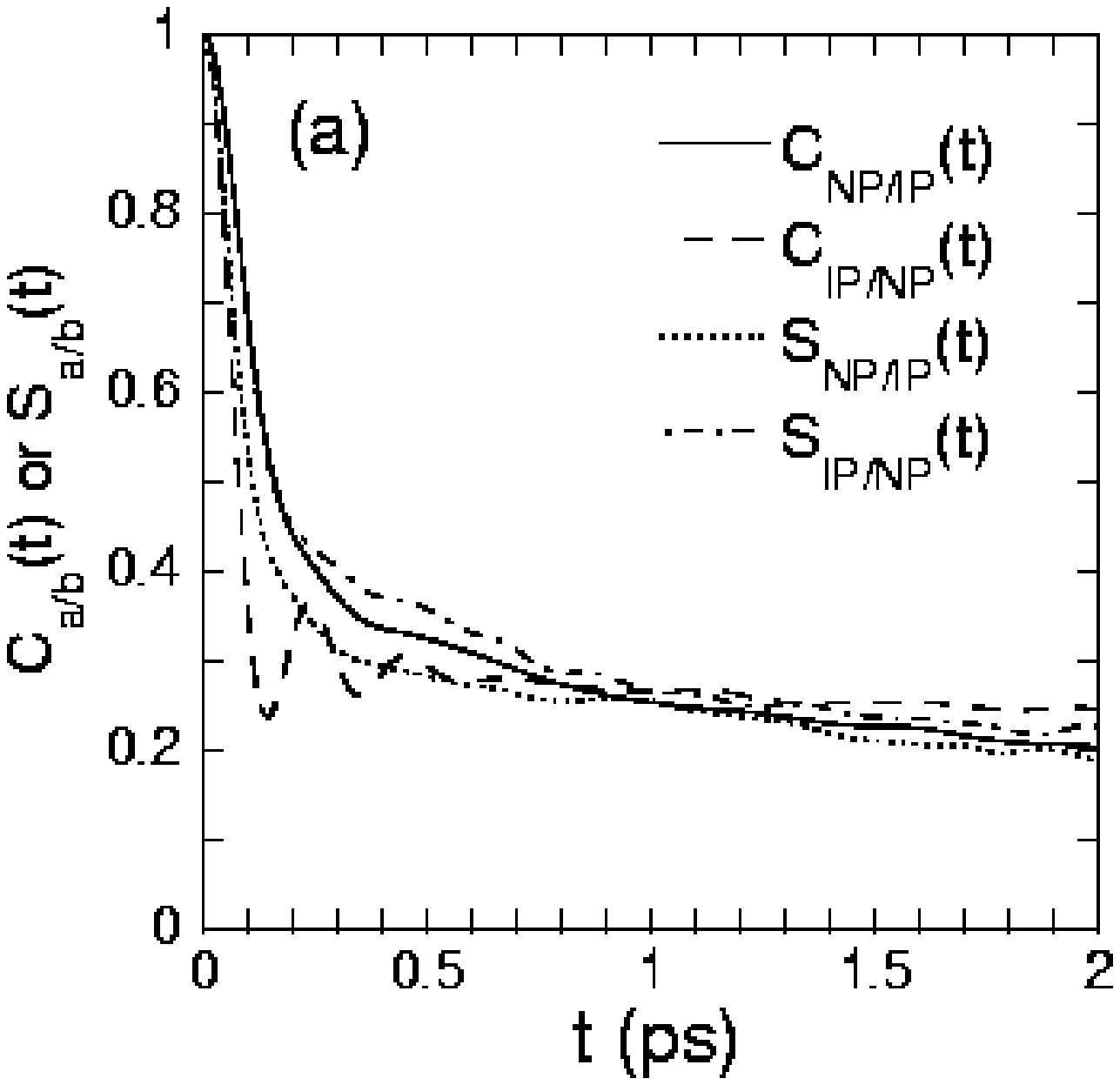}\vspace{0.3in}
\includegraphics[width=4.0in]{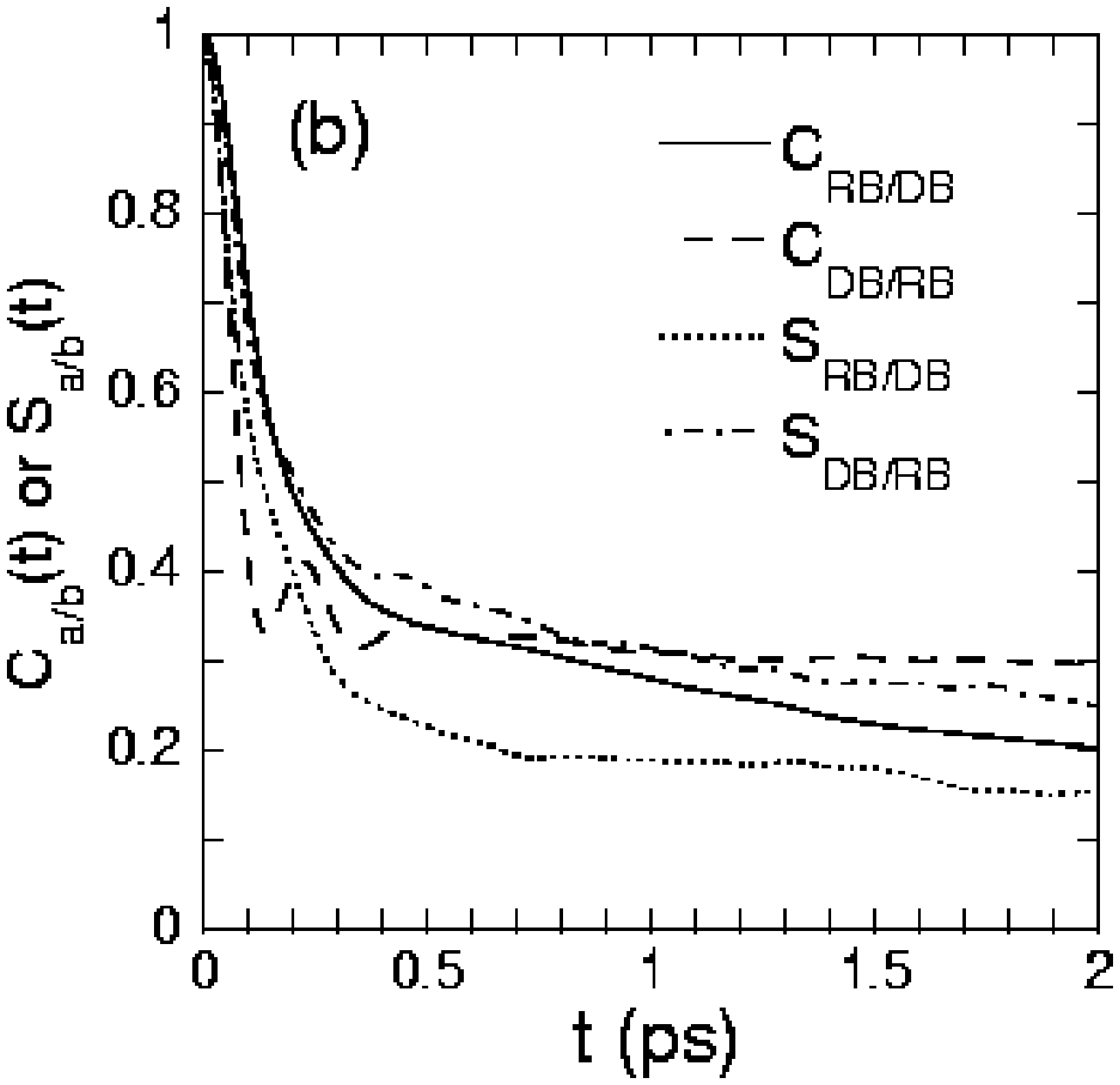}\vspace{0.3in}
\vspace{10cm}
\centerline{Fig.~\ref{fig:comp}}

\end{document}